\documentclass[10pt,a4paper]{article}
\usepackage{graphpap,epsfig,amssymb,graphicx,textcomp}
\begin{document}
\begin{center}
 \large{\bf Dynamical Phase transitions in Kuramoto model with distributed Sakaguchi phase}
\end{center}

\vskip 1cm

\begin{center}{\it Amitava Banerjee$^{1}$ and Muktish Acharyya$^{2}$}\\
\vskip 0.2 cm
{\it Department of Physics, Presidency University,}\\
{\it 86/1 College street, Kolkata-700073, INDIA}\\
\vskip 1 cm
{1. E-mail: amitava8196@gmail.com}\\
{2. E-mail: muktish.physics@presiuniv.ac.in}\end{center}

\begin{abstract}
In this numerical work we have systematically studied the dynamical phase transitions in the Kuramoto-Sakaguchi model of synchronizing phase oscillators controlled by disorder in the Sakaguchi phases. We find out the numerical steady state phase diagrams for quenched and annealed kinds of disorder in the Sakaguchi parameters using the conventional order parameter and other statistical quantities like strength of incoherence and discontinuity measures. We have also considered the correlation profile of the local order parameter fluctuations in the various identified phases. The phase diagrams for quenched disorder is qualitatively much different than those the global coupling regime. The order of various transitions are confirmed by a study of the distribution of the order parameter and its fourth order Binder's cumulant across the transition for an ensemble of initial distribution of phases. For annealed type of disorder, in contrast to the case with the quenched disorder, the system is almost insensitive to the amount of disorder. We also elucidate the role of chimeralike states in the synchronizing transition of the system and study the effect of disorder on these states. Finally, we seek justification of our results from simulations guided by the Ott-Antonsen ansatz. 
\end{abstract}

\vskip 1cm
\noindent {\bf I. INTRODUCTION}
\vskip 0.5 cm
Emergence of spontaneous synchronization is a recurrent theme in collective dynamics of coupled oscillators \cite{stro1a,stro1b,stro1c}. Nonlinear interactions among the oscillators can lock their amplitude or phases (or both) throughout the whole system or a part of it. This concept underlies many naturally observed phenomena, besides being present in systems of technological interest with an immense possibility to be exploited. Notable examples of synchronized collective rhythmic actions, which are mathematically modelled and analyzed at various levels of sophistication, include applause of an audience after a performance \cite{neda1a,neda1b,neda1c}, flashing of fireflies \cite{mat1,aria1}, chirping of crickets \cite{yeu1}, waving displays of fiddler crab claws during courtship \cite{bor1}, croaking of frogs \cite{aih1a,aih1b} and flagellar motion of algae \cite{bru1a,bru1b,bru1c,bru1d} from the animal world; and firing of coupled neurons \cite{mas1a,mas1b,mas1c,mas1d,mas1e,mas1f,mas1g,mas1h,mas1i,mas1j} and coordinated conduction of the heart's pacemaker cells \cite{mich1a,mich1b} in the human body to mention only a few. Spontaneous synchronization may have industrial and technological applications in systems like laser arrays \cite{des1a,des1b,des1c,des1d,des1e,des1f,des1g}, Josephson junction arrays \cite{vla1a,vla1b,vla1c}, microwave oscillators \cite{jung1a,jung1b,jung1c} and power grids \cite{adi1a,adi1b} among many. From a theorist's point of view, A large number of such systems arising from varied situations can be approached in a unified way through a simple yet paradigmatic model of Kuramoto. In the simplest form of this model, the dynamical variables are the phases $\theta_{i}(t)$ of $N$ coupled oscillators, whose dynamics is governed by a set of $N$ first-order differential equations (1) with sinusoidal coupling between the phase of each oscillator to those of its $R$ nearest neighbors at both sides \cite{kuramoto1,stro2,acebron},    
\begin{equation}
 \frac{d{\theta_{i}}(t)}{dt}= \omega-\frac{1}{2R}\sum_{j=i-R}^{i+R}\sin(\theta_{i}(t)-\theta_{j}(t)+\alpha)
\end{equation}
with $\omega$ being the natural frequency of each of the oscillators, which can be set to zero by a rotating frame transformation $\theta\mapsto \theta-\omega t$. The above model has $2$ parameters: the coupling radius, defined as $r=\frac{R}{N}$ and the Sakaguchi parameter $\alpha$, which is a phase lag in the coupling. In simulations, the periodic boundary identification $\theta _{N+p}=\theta _{p}$ is usually used. The Kuramoto model stands as the simplest yet the most successful understanding of synchronization processes till date, but, despite being thoroughly pursued, continues to host some unsolved mysteries \cite{stro2,acebron}.

To study the synchronization of natural processes using this model, one usually starts from an initial condition corresponding to a random distribution of oscillator phases. Over time, depending on the values of the coupling radius $r$ and the Sakaguchi phase $\alpha$, the system (1) demonstrates certain generic collective dynamical states for the phases $\{ \theta_{i} (t)\}$ \cite{acebron} :- (i) the asynchronous state, where the individual phase dynamics are uncorrelated  and which occurs for coupling radius smaller than a critical value, (ii) the globally synchronous state, where all the phases tend to assume a common value after a certain relaxation time and coherently oscillate together as a mass, and the (iii) chimera and multichimera states \cite{kuramoto2,panaggio}, which occur for certain non-local couplings and some particular values of the Sakaguchi phase, where the phases evolve as multiple synchronous groups, well-separated by incoherently moving oscillators. Each of these characteristic states offers many motivations to study them, both from theoretical and application-oriented perspectives. In particular, the relatively recent chimera and multichimera states are detected in coupled mechanical \cite{mar1a,mar1b,mar1c,mar1d,mar1e} and chemical oscillations \cite{panaggio,tin1a,tin1b,tin1c,tin1d} and SQUID metamaterials \cite{laz1a,laz1b,laz1c} and is thought to be related to the neural dynamics in the brain of certain birds and mammals during their unihemispheric slow-wave sleep \cite{panaggio,ma1a,ma1b,ma1c}, loss of coordination between human cardiac cells in ventricular fibrillation leading to cardiac deaths \cite{panaggio}, failure of power grids due to loss of synchronization between generators \cite{panaggio} and in the coexistence of laminar and turbulent patterns in the Couette flow \cite{dwi1}. Finally, we mention that the term 'chimera states' is usually attached to systems composed of identical oscillators \cite{panaggio}, which is not the case in this paper due to inhomogeneous Sakaguchi couplings. As is explained below, the coupling inhomogeneity naturally allows the formation these states. Thus, the states mentioned as `chimera' and `multichimera' are not so in the true sense. But if we neglect the coupling information, then these states are really not distinguishable from their conventional counterparts in terms of morphology, dynamics or statistical characterizations.

The Kuramoto model, showing such dramatic success over the time, has been appended with many modifications to make it more realistic. One of the obvious modifications is the addition of disorder in the form of random noise to mimic the behavior of realistic systems. This noise may arise, for example due to imperfect coupling between oscillators, presence of uncontrollable realistic parameters not taken into account in the model or environmental interference. A very general model containing multiple noise terms \cite{acebron,dai1a} is described by the set of equations:
\begin{equation}
\frac{d{\theta_{i}}(t)}{dt}= \omega_{i}+\sum_{(i,j)}K_{ij}\sin(\theta_{j}(t)-\theta_{i}(t)+\alpha _{ij})+\xi _{i}(t).
\end{equation}
Here $K_{ij}$, $\alpha _{ij}$, $\omega_{i}$ and $\xi _{i}$ are randomly distributed numbers withing suitable ranges. The model (2), along with some of its simplifications \cite{acebron} is quite thoroughly studied and is known to produce some interesting general features, usually absent in the original Kuramoto model \cite{acebron}. One of them is frustration arising from the fact that the coupling $\sum_{(i,j)}K_{ij}sin(\theta_{j}(t)-\theta_{i}(t)+\alpha _{ij})$ is non-vanishing for $\theta_{i}=\theta _{j}$ and consequently it is hard to find the configuration of the phases for which the couplings actually vanish. A related phenomenon is the existence of glassy states, where the oscillator phases remain randomly distributed but with their frequencies adjusting to each other and the accompanying slow relaxation dynamics. Quite a large amount of study is devoted to the conditions for emergence and microscopic dynamics of these states and their roles in synchronization \cite{acebron,dai1a,dai1b,dai1c,dai1d,dai1e,dai1f,dai1g,dai1h,dai1i,dai1j}. However, most of these studies are from the time before the discovery of the chimera states. So there is almost no work focused on the systematic investigation of macroscopic collective dynamics covering the whole regime of the parameter space of the model. A recent work \cite{chol1} has considered disorder in both the coupling strengths and coupling phases and two other works \cite{yi1a,yi1b}  have considered the effect of correlated noise in the system. They both have obtained some interesting dynamical features and have analyzed their bifurcation. Finally, some study has been done on a similar model of disorder as ours, but in the global coupling regime \cite{pazo1,pazo2,pazo3}. As we shall see below, most of our important results are from the non-locally coupled system. Hence, as will be clarified in the next section, qualitatively much different than the corresponding results from globally coupled system and we have characterized the order of the various dynamical phase transitions rigorously, which is absent in the last-mentioned study. So, in a sense, our work is complementary to theirs and somewhat completes the general understanding of the phase transitions of the model with distributed Sakaguchi phases. In summary, our model is much simpler than those involved in all the above works and we have investigated the role of the disorder in the phase and clarified how the chimera states are involved in various disorder-controlled phase transitions in the model. 

Here we have considered quenched and annealed disorder in one of the most important parameters in the model, namely, the Sakaguchi phase. In general, the microscopic dynamics and robustness of the chimera states under disorder is an interesting study in itself and some work has been done in this direction \cite{loo1a,loo1b,loo1c}, though a general study involving disorder in the Sakaguchi phase is still lacking. The role of Sakaguchi phase is very crucial in the Kuramoto model. For example, this parameter can be thought of as a time delay in the model equation \cite{panaggio,cro1} and its role is also important in setting a natural time scale in the model \cite{ban1}. So, a distribution in this parameter would imply a heterogeneity of microscopic time scales for various parts of the system, therefore the relaxation dynamics of the system to synchronized or chimera states would be affected. Furthermore, this parameter directly controls synchronization in the following way: for $\alpha <\frac{\pi}{2}$, the system spontaneously synchronizes for suitable parameter values as is seen numerically \cite{ban1}; for $\alpha=\frac{\pi}{2}$, the system is a completely integrable one and conservation of quantities like \cite{panaggio,wat1a,wat1b} $I=\Pi _{i=1}^{N} sin(\frac{\theta _{i}-\theta _{i+1}}{2})$ hinders synchronization from a random initial distribution of phases and for larger values of $\alpha$, the system never synchronizes for any parameter value, as is confirmed from numerical phase diagrams \cite{ban1}. In this sense, this parameter controls the balance between order and disorder in the model. Particularly in our case here, when there is an inhomogeneity in the Sakaguchi phase, some oscillator pairs have $\alpha<\frac{\pi}{2}$ between them, which favors synchronization and corresponds to an attractive coupling; while some pairs have the opposite case. This coexistence of attractive and repulsive coupling renders some of the oscillators locked in synchronization, while the others remain asynchronous. Therefore, chimera states are very natural in this set-up. However, as a contrasting case, it is also very interesting to study the dynamics when the Sakaguchi phase is time-independent but disordered spatially, due to the inhomogeneous coupling between individual oscillators; and also the case when it is disordered over time evolution of the system but spatially homogeneous. In the former case (quenched disorder), we have shown that for non-local coupling, the system has a rich phase diagram in the plane of mean and standard deviation of disorder involving both continuous and discontinuous synchronization transitions, with the nature of the transition being measured by how abruptly the system loses synchrony as the mean value of $\alpha$ gradually approaches $\frac{\pi }{2}$. The co-existence of both types of synchronization behavior vanishes for global coupling and there we only have discontinuous synchronization. This hints at the involvement of chimera states in the transition from synchronization to randomness as these states are usually seen in the cases with non-local coupling \cite{panaggio}. In this regard, we must mention that some works suggest that the type of synchronization (discontinuous/continuous) is dependent on the addition of inertial ($\ddot{ \theta _{i}}$) and/or noise terms \cite{bar1,sha1,kom1} to the conventional model. However, most of these works do not address or clarify the role of chimeralike states in the synchronizing transition, which we consider here. Also, the model we have taken is considerably simpler than those studied in these works.     

Understanding the nature of the collective dynamics with disorder will help in controlling systems of technological interest governed by such a dynamics. For example, if the couplings in laser arrays or Josephson junctions are experimentally controllable, then we can put disorder and control its strength there. We can change the nature of the synchronizing transition just by varying the amount of disorder. This will allow us to tune the degree of synchronization to an optimum level. This can also be used in the field of swarm robotics, where we can switch between the two modes of synchronization as per the needs while executing some complex operation. Applications of the Kuramoto-like models to robotics is already studied in the literature \cite{xu1a,xu1b,xu1c,xu1d} and a recent work \cite{mija1} indicates that a sensorial delay in motional responses to light signals (analogous to a phase lag) can largely affect the emergent behavior of phototactic agents.

 Disorder can also play a very important role in destroying unwanted synchronization and/or prevent its development in the system. In fact, it is easy to see that when the oscillators are identical ($\omega _{i}=\omega _{j}, \forall i,j$) and completely synchronized, then for all pairs $(i,j)$ of oscillators, $\theta _{i}=\theta _{j}$, and from Eq. $(2)$ (with $\xi _{i}=0$), the only way to destroy synchronization is sufficiently strong disorder in the Sakaguchi phase, under which global synchronization is no longer a steady state. This can be exploited to reverse the effect of synchronization, achieving which is the subject of an emerging field of research \cite{olt1}. One of the most dramatic and notorious examples of the negative impact of synchronization was the large lateral oscillation of the Millennium bridge of London \cite{stro10}. This occurred due to the rhythmic footsteps of the pedestrians walking on it, causing the newly-inaugurated bridge to be closed immediately to avoid possible damages and accidents. Desynchronization may also be important in treatments of epileptic seizures, which are believed to be caused due to excessive synchronization of neural activities in the brain which is often preceded by a significant desynchronization event as is evident from EEG and single-neuron signals \cite{jir1a,jir1b,jir1c,jir1d,jir1e}. The relevance of chimera states in Kuramoto-like model to epileptic seizures is already under active consideration, and control schemes are proposed to prevent such unwanted synchronization \cite{ral1a,ral1b,ral1c}. This is inspired by the observed correspondence between the spatiotemporal correlation profile of the EEG signals and the correlations among the coupled oscillators, when chimera states exhibit a dip in coherence just before they collapse to a globally synchronous state \cite{ral1a}. In general, disorder being inherent to any realistic physiological system, our present work may be relevant for the study of the onset and the processes leading to synchronization in the brain and may lead to proposals of control schemes to manipulate them. 

In this work, we have also studied the microscopic details of the dynamics of the chimeralike states under disorder. Controlling the chimera states is now an important theme in the study of dynamical systems \cite{sie1a,sie1b,sie1c}, particularly due to the proposed possibilities of technological, biochemical and computational applications of these states. In particular, the position of emergence of the synchronized region of the chimera state is highly uncertain and it depends on the initial conditions very sensitively \cite{kuramoto2}. So various control schemes are proposed so as to have that region present and stable at a pre-defined position of interest in the lattice. However, most of the control schemes proposed so far involve complicated themes like modulation of the coupling function between oscillators or the Sakaguchi phase with some controlled dynamical variable \cite{bick1} or even with the global order parameter itself \cite{sie1a,omel1}; spatial pinning \cite{gam1} and inclusion of localized excitable units \cite{ise1} in the system. In contrast, we have showed here that one can control the position of that region simply by putting inhomogeneity in the Sakaguchi phase. We show that only a well-controlled spatial inhomogeneity in $\alpha$ can squeeze or even split the synchronous region of a chimera state and position it at any arbitrary place on the lattice. This methodology may also be used to destroy unwanted synchronization at some region of the system, and thus may have clinical relevance in treatments of epilepsy. 

Finally, we have used the Ott-Antonsen ansatz for analysis of the dynamics of the order parameter for annealed disorder in the thermodynamic ($N \rightarrow \infty$) limit. We show that, for global coupling, a very simple single stochastic differential equation is shown to govern the dynamics of the system. The resulting phase diagram for the average asymptotic value of the order parameter matches qualitatively with the diagram that is obtained through direct simulation of the Kuramoto model. 

This paper is organized as follows. Section II describes the phase diagram of various collective dynamical states found with different values of the parameters related to quenched and annealed disorder. The states are distinguished using the conventional order parameter and two other statistical measures. We find that as we increase the standard deviation of disorder, the phase boundary corresponding to continuous transition ceases to be sharply defined and ultimately vanishes. Then we present a study related to the distribution of the order parameter and its fourth order Binder cumulant for the confirmation of the orders of the various transitions from the synchronized to the asynchronous state, as indicated by the phase diagrams. In section III, we are interested about the microscopic details of the dynamics under disorder and analyze the patterns in the local order parameter. The next section deals with the results of simulations related to the Ott-Antonsen ansatz and compares them to those obtained by direct simulation. Finally, section V summarizes the results and discusses future directions and possible extensions of the present work.

\vskip 1 cm

\noindent {\bf II. PHASE DIAGRAMS FOR QUENCHED AND ANNEALED DISORDER}
\vskip 0.5 cm
\noindent {\bf A. TRANSITIONS IN THE QUENCHED DISORDER}
\vskip 0.2 cm
In this work, we study the effect of disorder in the Sakaguchi phase of the Kuramoto model. The equations we take can be written in the general form
\begin{equation}
 \frac{d{\theta_{i}}(t)}{dt}=-\frac{1}{2R}\sum_{j=i-R}^{i+R}\sin(\theta_{i}(t)-\theta_{j}(t)+\alpha_{ij}(t))
\end{equation}
where the Sakaguchi phase is symmetric over any pair of oscillators, i.e., $\alpha _{ij}=\alpha_{ji}$ and is randomly selected from some well-known distribution. As the initial condition, we assume the phases $\theta _{i}(0)$ to be randomly and uniformly distributed in the range $[0,2\pi]$. We simulate Eq. (3) using fourth-order Runge-Kutta method with step size $dt=0.01$. After a large time, say, $1000$ units, when the system does not change much over time statistically, we wish to find out macroscopic properties of the evolved configuration. A much-used quantity is modulus of the complex order parameter $|Z|$ which measures the degree of coherence of the phases \cite{acebron}. For a given configuration of oscillator phases, this can be calculated from the following equation 
\begin{equation}
 Z(t)e^{i\psi }=\frac{1}{N} \sum_{j=1}^{j=N}e^{i\theta _{j}}
\end{equation}
where $\psi$ is the average phase of the oscillators. For the two limiting cases of completely asynchronous and globally synchronous states, $|Z|=0$ and $|Z|=1$ respectively; with intermediate values signifying partial synchronization of the oscillator phases.

As for the form of disorder, we first study quenched disorder, where the $\alpha $'s are time-independent and drawn, in our two cases of study, from normal and discrete bimodal distributions respectively. For the former case, the mean $m$ and standard deviation $d$ of the Gaussian distribution are the only two parameters in the model and the simulation of this distribution is performed using the well-known Box-Muller algorithm \cite{ross1}. Similarly, for the bimodal case, each of the $\alpha $'s can take a value either $m+d$ or $m-d$ with equal probability. With these, we obtain the phase diagram for the order parameter for global (Fig. 1) and non-local (Fig. 2) coupling over the parameter space of $m$ and $d$. The initial configuration of the system is random but same for all the $m$ and $d$ values in the throughout the phase diagrams. The plots clearly show the existence of three distinct phases in the system, two of them having two limiting values of $|Z|$ (corresponding to asynchronous and globally synchronous states) and one having an intermediate value of $|Z|$ (corresponding to partial synchronization). Furthermore, comparing the plots for two different distributions (parts a and b in Figs. 1 and 2), we notice that there is no significant qualitative or quantitative difference between them. So most of the properties of the system under disorder we shall address are insensitive to the exact form of the distribution of disorder we take. 

The most prominent feature evident from the plots is the loss of synchronization of the system beyond $m=\frac{\pi}{2}$, which we have already discussed in the previous section. However, for larger disorder dispersion, this loss of synchronization is seen to occur for smaller values of the mean disorder, so the boundary separating the synchronous and asynchronous phases is not exactly vertical in the $m-d$ plane. However, a comparison of Fig. 1 and Fig. 2 shows that for global coupling, the system attains synchronization through a single phase boundary for any value of the disorder dispersion. But in the non-local coupling case, it does so in two different ways. For lower values of disorder the synchronizing transition is a two-step process. The system first undergoes a transition to a state with an intermediate phase ordering (having the asymptotic value of $|Z|$ significantly less than 1), from which it again goes to asynchrony through another phase boundary as $m$ increases. Thus we have two closely-spaced phase boundaries, both bounding the intermediate state. These two boundaries seem to merge together for higher values of disorder and there we only see a single well-defined phase boundary.

We next proceed to detect the order of various transitions involved by noting the change of order parameter across the phase boundary in two different parameter regimes. We obtain the histogram (Fig. 3) of the asymptotic values of the order parameter for a fixed realization of Gaussian quenched disorder and a large number of random initial conditions of the phases with parameters near the phase boundaries, which is a standard technique to detect the order of the phase transition \cite{bind1a,bind1b} and has been used in the present context also \cite{sha1}. In any of the synchronous, intermediate and asynchronous phases, sufficiently far from the transition regions, the histogram is seen to be unimodal and sharply peaked, respectively, at a larger (near $1$), intermediate (around $0.7$ in our case) and smaller (near $0$) values of $|Z|$ in the three cases. This shows that the system is monostable at parameter values away from the phase boundaries. We thus demonstrate that even if we plotted the phase diagrams (figs 1 and 2) using one single initial condition, they are representative plots for any typical random initial configuration of the system. However, near the transition boundary, the system becomes bistable and so the histogram shows a nonzero spread. For continuous transitions, there is supposed to be a parameter range near the transition region where the histogram is significantly spread over the whole set of possible values of $|Z|$ attainable for that parameter range, corresponding to systems with intermediate amounts of coherence. Existence of this histogram profile in Fig. 3a. shows that the first transition between synchronous to partially-ordered states for lower disorder in the non-local coupling case is a continuous one. However, during this transition, the order parameter varies only little, between 0.7 to 1. Thus the histogram is visually not a very conclusive evidence of continuous transition due to the fact that the profile of the histogram in the relatively small range of the variable is hard to identify due to fluctuations. So, below we shall use two more direct and sophisticated ways to find the nature of the transition. In contrast to this case, for abrupt transitions, there should be no such parameter regime at which this large spread of histogram happens. Rather, throughout the transition regime, we should get a bimodal distribution with two very sharp peaks at the two $|Z|$ values representing the two phases across the phase boundary, which reveals the absence of any intermediate value of the order parameter. This is indeed the case for the transitions between partially ordered and asynchronous states for both smaller and larger disorder values and non-local coupling (Figs 3b and 3c). 

For a more direct investigation, especially of the continuous transition, we also plot the average and standard deviation of the order parameter in Figs 4a and 4a$^{/}$ Supporting our expectations from the profile of histograms, we do see continuous/abrupt changes of the mean order parameter across the relevant transition points as we expected. We also note a rise of standard deviation of the order parameter near the expected transition regions, which is due to bistability of the system near phase boundaries. Furthermore, at high disorder strength, we find the transition to the incoherent state is noisy and the mean of the order parameter is of the same order as the standard deviation (Fig. 4a$^{/}$). 

Finally, we use the Binder's fourth order cumulant $U_{4}(m,d)$ to probe the nature of the transition. This quantity is defined by the equation
\begin{equation}
 U_{4}(m,d)=1-\frac{\left<|Z|^{4}\right>}{3\left<|Z|^{2}\right>^{2}}
\end{equation}
where the brackets denote averaging over the ensemble of initial configurations. After its introduction and development \cite{bind1a} it has turned out to be a widely used tool to identify the nature of phase transitions in a variety of circumstances ranging from model magnetic systems \cite{bind1b,muk1a,muk1b,muk1c}, liquid crystals \cite{gho1}, foams \cite{has1}, flocking systems \cite{bhat1} and opinion dynamics \cite{mu1} to quantum chromodynamics \cite{cza1}. This quantity $U_{4}$ detects the order of a phase transition in the following way \cite{bind1a,bind1b}:  (i) In a phase with a nonzero order parameter value, but with very small order parameter fluctuation, $\left<|Z|^{4}\right>\approx\left<|Z|^{2}\right>^{2} $ and so it attains the value $\frac{2}{3}$; while in a completely disordered phase, it takes the value 0 and (ii) during a first order (discontinuous transition) transition, it shows a sharp jump to a very sharp minimum of a large negative value while for a second order (discontinuous) transition, it decreases from $\frac{2}{3}$ but remains positive. In our case, it shows a sharp negative excursion at $m=1.535, d=0.2$ (Fig. 4b) and multiple large negative excursions for $d=0.85$ (Fig. 4b$^{/}$), confirming discontinuous transition. Finally, it also exhibits a small dip around $m=1.44,d=0.2$ (Fig. 4b), which clearly indicates a continuous phase transition as we anticipated.

As we noted before, the two-step synchronization process appears only for non-local coupling, which is the widely-accepted condition for chimera states also \cite{panaggio}. In fact, as we show below, for continuous transition, the intermediate values of the order parameter correspond to chimeralike states, whose formation changes the order parameter continuously. To identify these states, we use some specialized statistical measures, namely, the strength of incoherence and discontinuity measures, both of which are introduced, developed and used in earlier works \cite{gopal1,gopal2}. To calculate them, we define $z_i=\theta _{i}-\theta _{i+1}$ the local fluctuation of phases, which is negligible in a ordered oscillator cluster and assumes a large value for incoherent oscillators. Then we divide $N$ oscillators into $N_g$ groups of equal length side-by-side and calculate the time-averaged
variance of $z_i$ in the k-th group defined by the equation:
\begin{equation}
\sigma_k= \left\langle \sqrt{{\sum_{j=1}^{N_b}(z_i^k-{\bar z_i^k})^2} \over {N_b}}\right\rangle_{t}
\end{equation}
measuring the fluctuation of phases in a single group. Next we define a new variable $\lambda _{k}$ for each group by the equation
\begin{equation}
\lambda_k=\Theta (\delta - \sigma_k)
\end{equation}
where $\delta$ is a sufficiently small value taken as the threshold of phase dispersion in each group. $\Theta$ is the Heavyside step
function, which selects only the groups which have a sufficiently large phase coherence so that the dispersion $\sigma _{k}$ is smaller than the prescribed threshold $\delta $.  Finally, we define the strength of incoherence $S$ as \cite{gopal1,gopal2}
\begin{equation}
S = 1 - {{\sum_{k=1}^{k=N_g} \lambda_k} \over {N_g}}.
\end{equation}
Here, for complete phase synchrony $S=0$, for complete randomness $S=1$ and $0<S<1$ implies chimera or multichimera states. However, this measure fails to distinguish well between chimera and multichimera states.
This is achieved through the discontinuity measure defined as \cite{gopal1,gopal2}
\begin{equation}
\eta = \frac{\left(\sum_{k=1}^{k=N_g} |\lambda_k - \lambda_{k+1}|\right)}{2}
\end{equation}
with the periodic boundary condition $\lambda_{N_{g}+1}=\lambda_{1}$. 
$\eta=0$ represents both completely synchronous and completely asynchronous phases, $\eta=1$ implies single chimera states and larger values of  $\eta $ imply multichimera states. With these detailed techniques for characterization of dynamical states, we now obtain the phase diagrams for strength of incoherence (SI) and discontinuity measure (DM) in Figs. 5-7.

In the phase diagrams of SI and DM, for the case of non-local coupling (Fig. 5), a broad band of intermediate values of SI and DM exist between the phase boundaries at $m=1.4$ and $m=1.56$ for $d=0 - 0.2$. They correspond to chimeralike and multichimeralike states (Fig. 5), which we hypothesized earlier and which are absent in the global coupling case (Fig. 7). Moreover, as we shall show in the next section, for non-local coupling, an increase of $m$ towards $\frac{\pi }{2}$ implies a decrease in the size of the coherent region of chimeralike states and a loss of their ordering in time but keeping $|Z|$ at a practically constant value, so the system desynchronizes only abruptly. A comparison of the phase plots for SI, DM and those for the order parameter in the case of non-local coupling (Fig. 2) shows that in most of the parts they match, except for some regions in the top left corner of the figures, where we found states with clustered oscillators. These states also occur at the bands near the phase boundary at $d=0.3$ in both global and non-local coupling. They are characterized by SI and DM values similar to those for chimeralike or asynchronous states; but unlike chimera states, these do not exhibit any significant decrease of the order parameter from $1$. However, as we have discussed above, the transition of these states to asynchrony with continuously increasing $m$ is non-monotonic and highly discontinuous. These cluster states also occur for the case of global coupling where we see intermediate values of SI and DM for some parameter regimes but there are generally no chimeralike states as we have checked. 
Taking account of all the above information furnished by $Z$, SI and DM, we produce a schematic phase diagram for non-local coupling in Fig. 6. To the best of our knowledge, except near phase boundaries, the system always attains the states mentioned in the schematic, as is evident from the unimodal histrograms in Fig. 3 corresponding to $m$ and $d$ values sufficiently far from the phase boundaries. 
\vskip 0.5 cm
\noindent {\bf B. ANNEALED DISORDER}
\vskip 0.2 cm

We now turn our attention to the case where the disorder in $\alpha$ is spatially homogeneous, but fluctuating in time instead. Here we assume that the Sakaguchi phase is chosen independently at each time step, from the uniform distribution in the range $[m-d,m+d]$, where $m$ and $d$ are the mean and maximum deviation of the disorder. As before, we do the phase diagram for the parameter space of $m$ and $d$ for both non-local and global coupling (Fig. 8). Interestingly, unlike the case for quenched disorder, here disorder has little effect on the dynamics, as is evident from the phase diagrams. The phase boundary is near $m=\frac{\pi }{2}$ as expected and is almost vertical. For global coupling, the transition to asynchrony is sharp. For non-local coupling, there is an involvement of chimeralike states, hence a band of intermediate values of $|Z|$ is present for $m$ in the range $1.45-1.55$. This is supported by DM and SI phase diagrams showing existence of such states (Fig. 9). Unlike the case of quenched dynamics, here the chimeralike state can tolerate much large values of disorder. In fact, as we shall show in the next section, chimeralike states exhibit a collective motion of the synchronized part of the system, but maintain a similar value of the order parameter as a whole.

\vskip 1 cm
\noindent {\bf III. MICROSCOPIC DETAILS OF DYNAMICS UNDER DISORDER}
\vskip 0.5 cm
In this section, we shall investigate the effect of disorder on microscopic details of the dynamics and probe the changes in the configuration of the oscillator phases when the system undergoes various transitions. Firstly, we show how one can manipulate the chimeralike state by putting a well-planned profile of the spatial heterogeneity of $\alpha $ in the system. In this respect, it is useful to recall that synchronization is impossible for $\alpha >\frac{\pi }{2}$ and so, the synchronous portion in a chimeralike state avoids those regions in the lattice where coupling between oscillators has this condition. To exploit this, we first start with the usual condition for a chimeralike state (say, $\alpha _{ij}=1.46 \forall i,j$) and let the chimera state form. Next, we suddenly change the $\alpha$ to the value of, say, $2$ for the pairs of oscillators which are in the synchronized part, to which the system responds in two possible ways:- 

\textbf{(a)} If there is sufficient space at both sides of the previously synchronized region containing oscillator pairs with $\alpha=1.46 $, but the other oscillator pairs beyond the region has $\alpha=2$ again, then the synchronized part splits into two and moves to the two sides of the previous region (Fig. 10a). The rest of the oscillators continue to be asynchronous like before, so this is a multichimeralike state. The phases of the oscillators in these two synchronous parts now have much lesser frequency than the same in the original chimeralike state. However, this splitting is reversible. If we again set $\alpha =1.46$ globally, we again have a chimeralike state similar to the original in the frequency of the synchronous phases, but the synchronous region is now situated at a different part in the lattice.

\textbf{(b)} If there is sufficient space at only one side of the originally synchronous region, and the other side has oscillator pairs with $\alpha =2$, then the synchronous part drifts to the former side till it meets oscillators with $\alpha =2$ again (Fig. 10b). This change is reversible as well, we can make the synchronous part drift again to its former position in lattice using the above procedure.

On the other hand, if we do not change the Sakaguchi phase of the synchronized oscillators, but put $\alpha =2$ for the oscillators at both sides of it, then the synchronous part is trapped. Next, we let some of the oscillator pairs at the side of the synchronous part have $\alpha =2$ also so that they lose synchrony and, as a result, the synchronous part decreases in length (Fig. 10c). This is a reversible change also and we restore the length of the synchronized part by setting $\alpha =1.46$ globally. In each of the three cases above, the position, as well as the extent of the region containing the oscillators having $\alpha =2$ in their coupling has to be chosen carefully. If the synchronous part of a chimeralike state does not get sufficient space in the lattice, the chimeralike state itself is destroyed and travelling waves are formed. These observations are useful in engineering chimeralike states in physical systems like laser arrays where the coupling is easy to control.

We now turn our attention to the effect of quenched Gaussian disorder on chimeralike states and see how the system loses order with increasing $m$ in a discontinuous way. We note that, in the lower values of $d$, as the mean value of the Sakaguchi phase nears $\frac{\pi }{2}$, the synchronous part of the chimeralike state begins to move along the lattice and its length fluctuates in an unsteady manner (Fig. 11). However, this is not reflected in the global order parameter which does not change significantly during this. This leads us to conclude that the seemingly asynchronous part of a chimeralike state also contains significant phase ordering. Finally, the synchronous part vanishes and the system becomes totally asynchronous. For higher values of $d$, there is no chimeralike state, and cluster states give way to asynchrony in a discontinuous manner.

When our system has non-local coupling, its dynamics can exhibit interesting spatial patterns and correlations at various length scales, which are neglected in global measures like the global order parameter, SI or DM. To probe and statistically characterize such dynamic patterns, we study the time-averaged spatial autocorrelation of the local order parameter. We define the local complex order parameter $Z(i)$ by the following equation
\begin{equation}
 Z(i)=\frac{\sum_{j=i-R_{l}}^{i+R_{l}} e^{i\theta_{j}}}{2R_{l}+1}
\end{equation}
so that its amplitude measures the ordering of the oscillator phases in only a small neighborhood of size $2R_{l}$ around a lattice site $i$. the size of the neighborhood $2R_{l}$ should be carefully chosen so that it is sufficiently large to smoothen the random phase fluctuations varying irregularly from site to site, while still being sufficiently small compared to the lattice size so that the information of spatial inhomogeneity is not lost. Then the long-range correlation pattern of the phases can be captured in the autocorrelation of the fluctuation of the local order parameter, defined by the equation
\begin{equation}
 A'(p)=\left<(|Z(i)|-\left<|Z|\right>_{i})(|Z(i+p)|-\left<|Z|\right>_{i})\right>_{i,t}
\end{equation}
 and normalized so as to have a maximum value of 1,
\begin{equation}
 A(p)=\frac{A'(p)}{A'(0)}.
\end{equation}
In Fig. 12 the correlation profile is plotted for various values of $m$ with low and high disorder dispersion. As we note, in both cases the correlation profile shows a clear visual distinction between the collective dynamical states we discussed. In particular, in Fig. 12a, the chimeralike state shows a significantly much longer correlation length than other phases. The state is also seen to exhibit a particular length scale of the order of the size of the synchronous part of the chimeralike state, above which $A(p)$ is negative and below which it is positive. This occurs due to the fact that for length scales smaller than this, both the lattice sites $i$ and $i+p$ in Eq. (11) either belongs to the synchronous part (in which case both of them have greater value of the local order parameter than the global mean $\left<|Z|\right>_{i}$) or the asynchronous part (in which case both of them have lesser value of the local order parameter than the global mean) so that the product $(|Z(i)|-\left<|Z|\right>_{i})(|Z(i+p)|-\left<|Z|\right>_{i})$ in Eq.(11) is positive and vice versa. Apart from this, the synchronized and asynchronous states also have their own correlation profiles, which do not seem to have such prominent length scales. However, as is obvious from Fig. 12b the profiles for the asynchronous states contain much less contrast than the ordered cluster states at high disorder strength. 
 
For the case of annealed disorder, the phase diagrams reveal that there is no appreciable effect of disorder, but the response of the system near $\alpha =\frac{\pi}{2}$ is interesting. Initially the synchronized portion in a chimeralike state is almost stationary in the lattice (Fig. 13a). As $\alpha $ increases, that portion begins to move in an apparently random manner across the lattice but keeping its size almost intact, till we can no longer distinguish that part separately (Fig. 13b and 13c). 
 
\vskip 1 cm

\noindent {\bf IV. ANALYSIS USING THE OTT-ANTONSEN ANSATZ}
\vskip 0.5 cm  
The dynamics of the continuum limit ($N \rightarrow \infty $) of the Kuramoto model can be studied using the Ott-Antonsen ansatz, which describes the dynamics of the Kuramoto order parameter by only a very small number of differential equations \cite{ott1}. In this work, We shall use the technique to study the case for annealed disorder only, the case for quenched disorder can be studied later. The main problem in the latter case is that the coupled random variable $\alpha _{ij}$ or $\alpha (x,x')$ in the system is not a separable function of its arguments in our work, which is the case for most of the studies done so far in analyzing Ott-Antonsen models for disordered Kuramoto systems \cite{chol1,pazo1,pazo2,pazo3,olt1,lop1a,lop1b}. So, we proceed with the case of annealed disorder. Non-local Ott-Antonsen equations used below have been derived, analyzed and used in previous studies \cite{ban1,nlott1,nlott2}. In the thermodynamic limit, one can define a distribution function $f(x,\omega,\theta,t)$ of the oscillator phases, which, by the proposal of Ott and Antonsen \cite{ott1}, has the Fourier components given by various powers of a (still undetermined) function $a(x,\omega ,t)$, i.e.,
\begin{equation}
f(x,\omega,\theta,t)=\frac{g(\omega)}{2\pi}\left(1+\sum_{n=1}^{\infty}[a^{n}e^{in\theta}+(a^{*})^{n}e^{-in\theta}]\right).
\end{equation}
This, alongwith our initial assumption of identical natural frequencies set to $0$ in a rotating frame (i.e., $g(\omega )=\delta(\omega -\omega _{0}),\omega _{0}=0$) gives the two differential equations for the spatio-temporal dynamics of the local order parameter:-
\begin{equation}
Z(x,t)=\frac{1}{2R}\int_{x-R}^{x+R}a^{*}(x',t)dx',
\end{equation}
and
\begin{equation}
\frac{\partial a}{\partial t}=\frac{1}{2}[Z^{*}e^{i\alpha(t) }-a^{2}Ze^{-i\alpha (t)}]
\end{equation}
where $a=a(x,0,t)$ is now a function of space and time. At this point, we use the specific form of $\alpha (t)$ which we used in our numerical simulations, i.e., the time series of $\alpha $ is a collection of random numbers chosen from an uniform distribution between $m+d$ and $m-d$ and generated independently at each time step of simulation. 

For the case of global coupling, we can further simplify Eqs. (14) and (15). In this case, the integration in Eq. (14) runs over the whole lattice, so the integral $Z$, and consequently $a$ (by Eq. (15)) is independent of $x$. So, Eq. (14) gives $Z=a^{*}$ and substitution of this to Eq. (15) yields the dynamics of the global order parameter
\begin{equation}
 \frac{dZ}{dt}=\frac{1}{2}[Ze^{-i\alpha (t)}-|Z|^{2}Ze^{i\alpha (t)}].
\end{equation}
We are interested mainly about the amplitude of the order parameter, so we define $Z=\rho e^{i\lambda }$, where $\rho$ and $\lambda $ are real. Equating real and complex parts of Eq. (16) separately, we obtain the final set of equations
\begin{equation}
 \frac{d\rho }{dt}=\frac{1}{2}(\rho -\rho ^{3})\cos\alpha (t) 
\end{equation}
and
\begin{equation}
 \frac{d\lambda }{dt}=-\frac{1}{2}(1+\rho ^{2})\sin\alpha (t). 
\end{equation} 
When $d=0,\alpha =m$, Eq. (17) has only $2$ fixed points: $\rho =0$, the asynchronous phase and $\rho =1$, the synchronous phase. The latter is a stable fixed point if and only if $m<\frac{\pi }{2}$, a limit about which we discussed earlier. In both of the cases, asymptotically $Z$ evolves in time via only a phase oscillation with a time scale of the order $\frac{\pi }{\sin m}$, which we identified in our earlier work \cite{ban1}. However, returning to the general case, we numerically solve Eq. (17) using Eulerian discretization with step size $dt=0.01$ for various values of $m$ and $d$ and plot the asymptotic values of $\rho $ in Fig. 14. In our simulations of Kuramoto equations, we started from configurations involving uniformly distributed phases, which correspond to $\rho =0$. However, $\rho =0$, being a fixed point, can not be an initial condition to solve equation. So, we choose a small non-zero value, $\rho=0.001$ as the starting condition. As another implication of the fact that $m=\frac{\pi }{2}$ is a transition region, we also note that the time taken by $\rho$ to reach a saturation value increases as the value of $m$ nears $\frac{\pi }{2}$. Finally, we see that the phase diagram (Fig. 14) conforms well to that (Fig. 8a) found by direct simulations of the model (3). In particular, there is little effect of disorder on the dynamics and the phase boundary at $m=\frac{\pi }{2}$ is almost vertical.

\vskip 1 cm
\noindent {\bf V. DISCUSSION AND CONCLUSIONS}
\vskip 0.5 cm
In this work, we systematically studied various effects of inclusion of disorder in the Sakaguchi phase of the Kuramoto model. In this case, as we have shown here, one can control the nature of synchronization by tuning the amount of disorder in the system, which is not the case for annealed disorder. The nature of the synchronizing transitions are established here by studying the distribution of order parameter and its fourth order Binder's cumulant in the vicinity of transition point. In the annealed disorder case, the simulation results are shown to be excellently in tune with the predictions of the corresponding Ott-Antonsen low-dimensional formulation of the system. The coexistence of both continuous and discontinuous transitions may inspire one to associate thermodynamic properties and quantities to the model. In particular, the discontinuous transition may be accompanied by a latent heat arising due to a difference in entropy (properly defined) of the synchronous (more ordered) and asynchronous (less ordered) states. This may be modeled by the addition of a $\rho ^{5}$ term in Eq. (17), which then becomes the magnetization dynamics of a ferromagnetic sample with a Landau-type free energy having a term proportional to the sixth power of the order parameter \cite{chaikin1} exhibiting a tricritical behavior of the magnetic sample. However, it should be justified whether this type of equation actually holds for quenched disorder at all.

We have also investigated the role of chimeralike states in this noise-induced transition. Interestingly, chimeralike states, whenever formed, are surprisingly stable under disorder and undergo interesting changes in their structure and dynamics. In the dynamic changes, we have seen that the synchronous part of a chimeralike state oscillates randomly about some mean position, move through the entire lattice apparently erratically, change its length over time or even break apart. This inspires us to formulate the dynamics of a chimeralike state under disorder using only few dynamical coordinates like mean position and size of the synchronized region and the common phase of the oscillators situated there, rather than resorting to descriptions of the field $\theta (x,t)$ or the distribution function $f(x,\theta ,\omega ,t)$. A better understanding of all these dynamical effects of disorder on chimeralike states will help us achieve a control scheme of chimeras through disorder, which may be helpful in both technology or medicine. In particular, the destruction of the synchronous part by sufficiently strong disorder is analogous to disorder-driven melting \cite{nelson1}. Furthermore, one can study whether there is any universal response of chimera states in presence of noise which is independent of the exact model in which the chimera comes in. In conclusion, we wish to remark that the inclusion of noise in one of the most important parameters of the Kuramoto model to mimic natural dynamics leads to a rich spectrum of both microscopic and macroscopic phenomena.   
 
\vskip 1 cm
\begin{center}
-----------------------------------------------------------------------------------------------------------------------
\end{center}

\newpage
\begin{center}
 \textbf{\underline{Figures}}
\end{center}

\begin{figure}[h]
\begin{center}
\begin{tabular}{c}
        \textbf{(a)} \resizebox{8cm}{!}{\includegraphics[angle=-90]{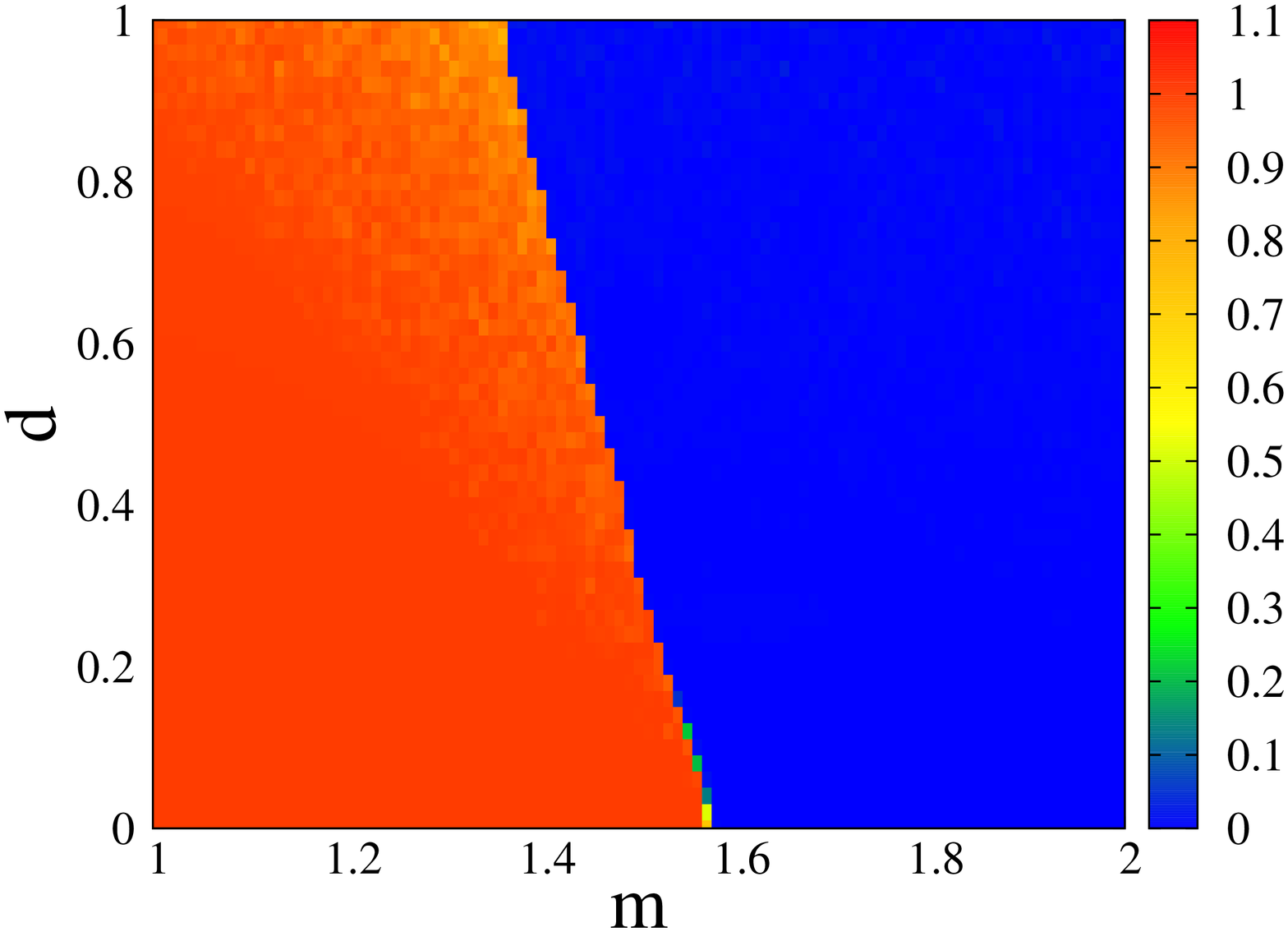}}
         \textbf{(b)}\resizebox{8cm}{!}{\includegraphics[angle=-90]{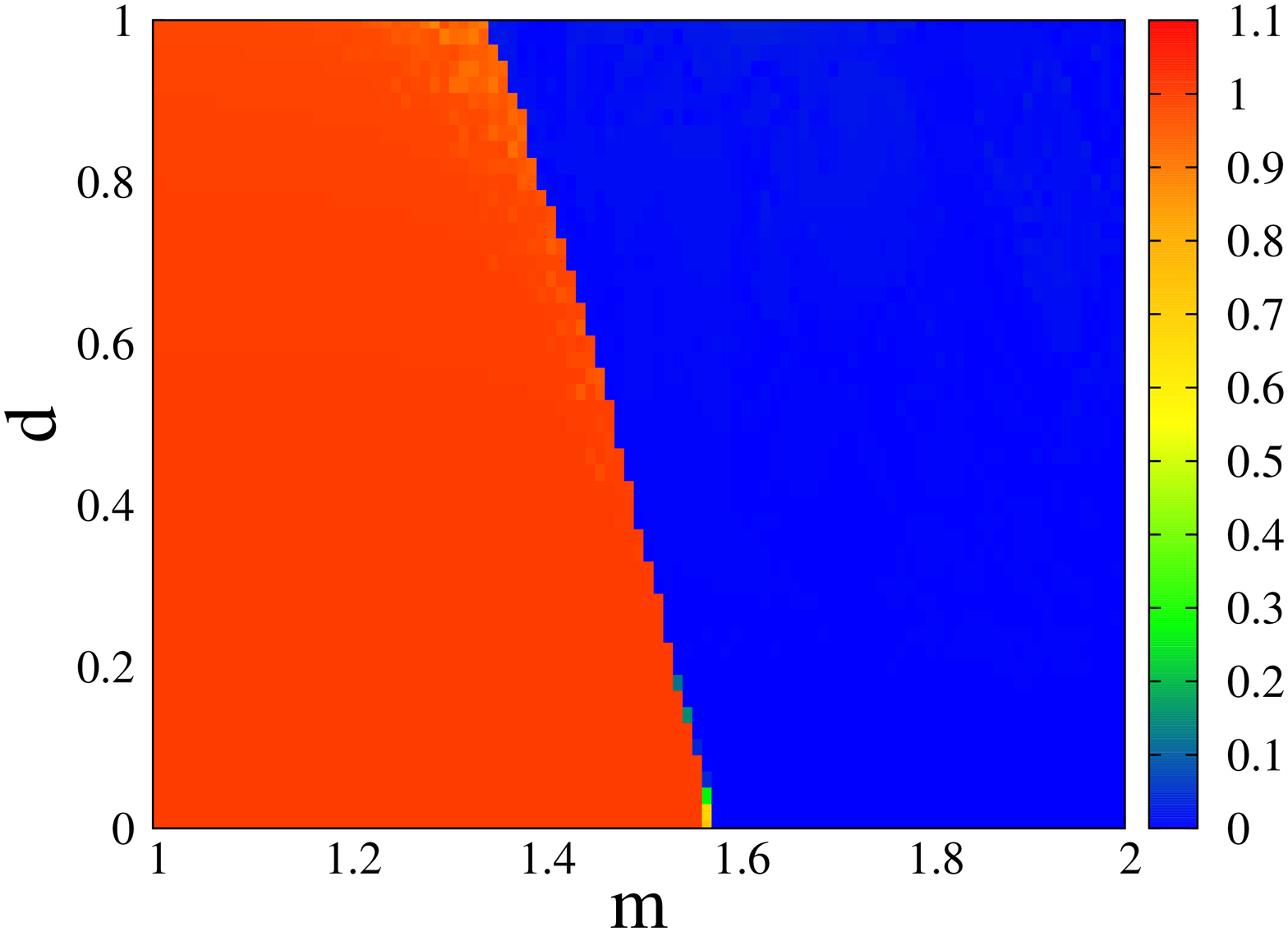}}
        
                  \end{tabular}
 \caption{{(Color Online) Phase Diagram for modulus of the order parameter $|Z|$ at time 1000 unit for (a) Gaussian and (b) Discrete Bimodal quenched disorder, with $N=100, r=0.49$, with identical initial random configuration of phases showing only discontinuous synchronization.}}
\end{center}
\end{figure}
\vskip 2 cm
\begin{figure}[h]
\begin{center}
\begin{tabular}{c}
        \textbf{(a)} \resizebox{8cm}{!}{\includegraphics[angle=-90]{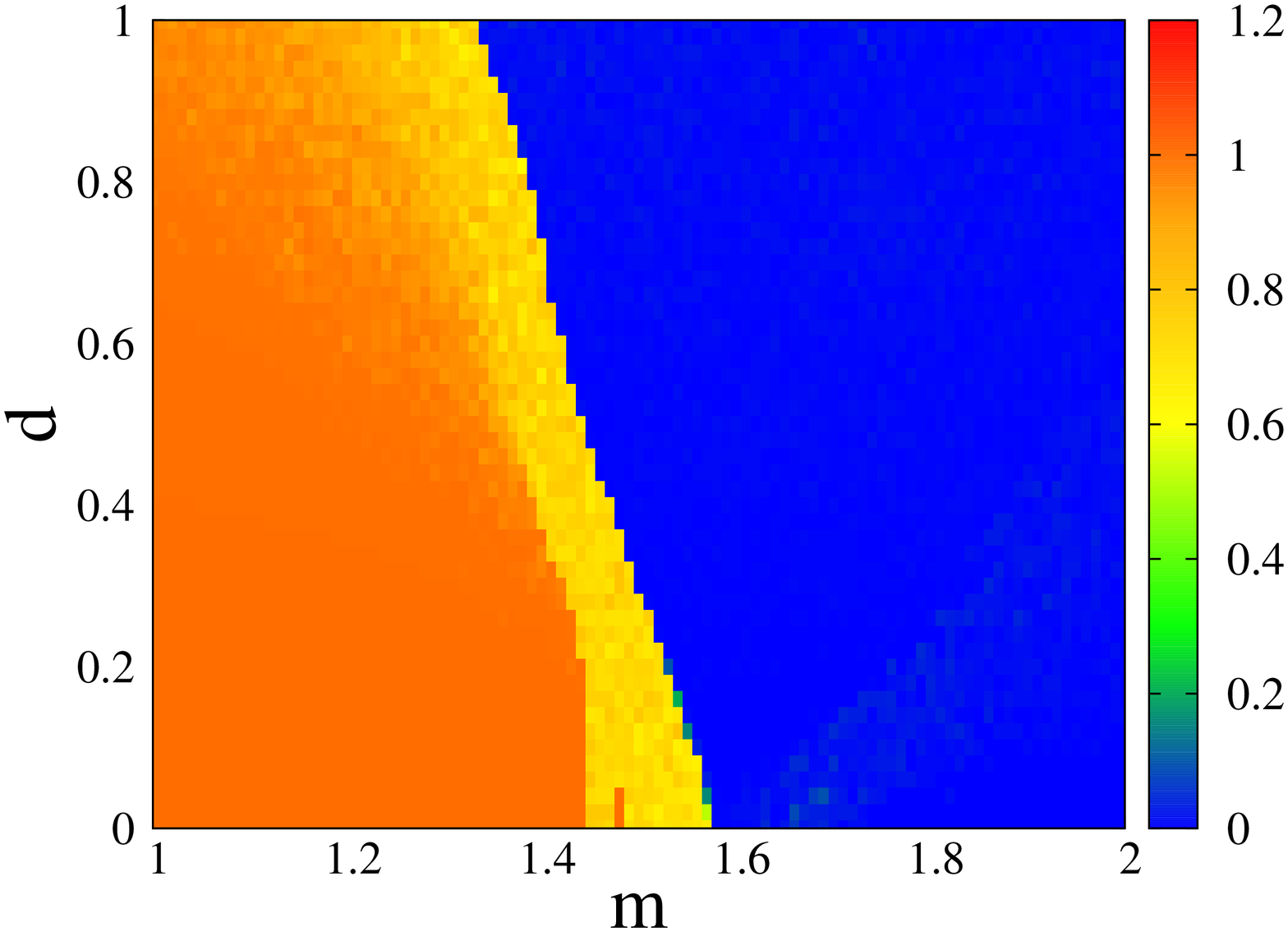}}
         \textbf{(b)}\resizebox{8cm}{!}{\includegraphics[angle=-90]{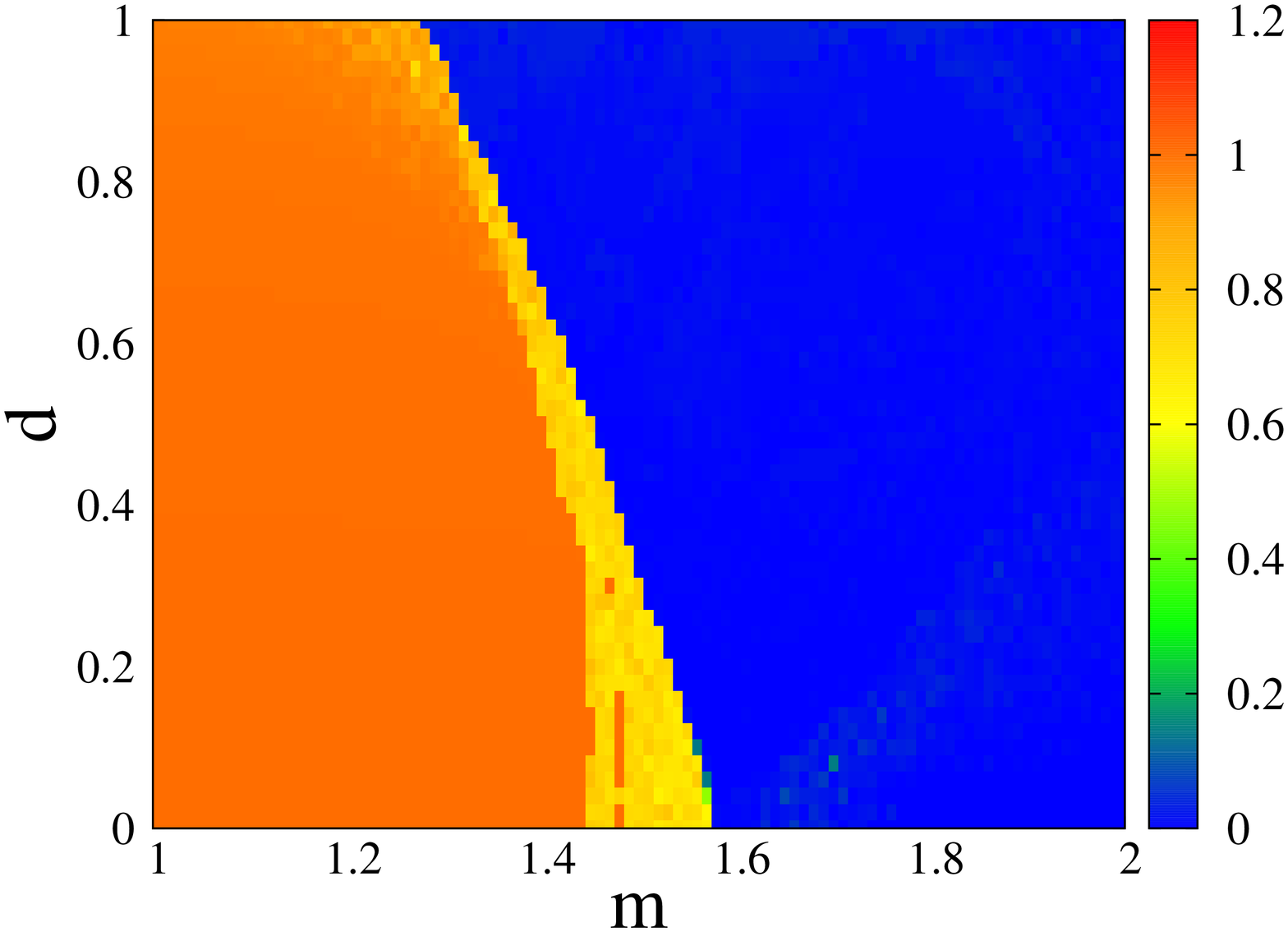}}
        
                  \end{tabular}
 \caption{{(Color Online) Phase Diagram for modulus of the order parameter $|Z|$ at time 1000 unit for (a) Gaussian and (b) Discrete Bimodal quenched disorder, with $N=100, r=0.35$ , with identical initial random configuration of phases.}}
\end{center}
\end{figure}

\newpage
\begin{figure}[h]
\begin{center}
\begin{tabular}{c}
        \textbf{(a)} \resizebox{8cm}{!}{\includegraphics[angle=-90]{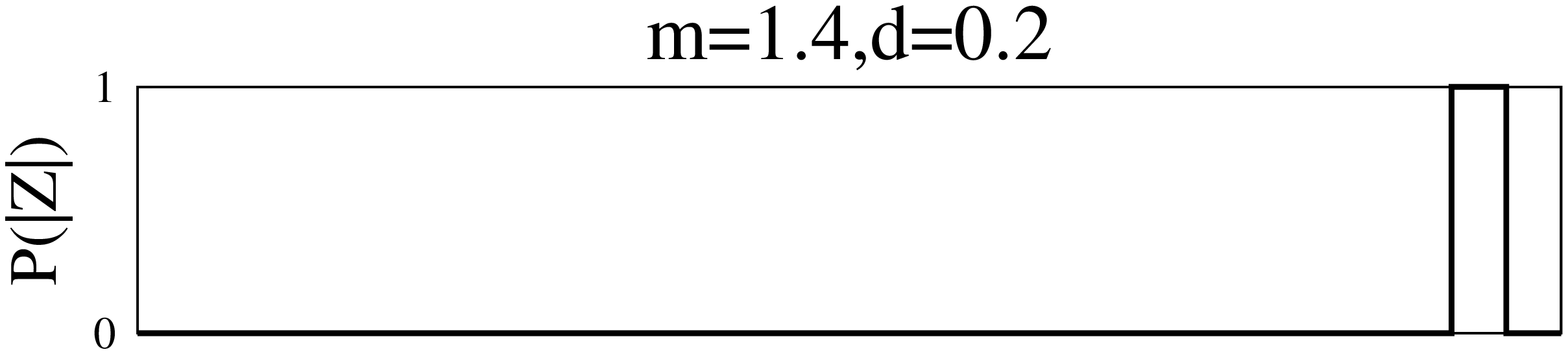}}\\
        \hskip 0.5 cm\resizebox{8cm}{!}{\includegraphics[angle=-90]{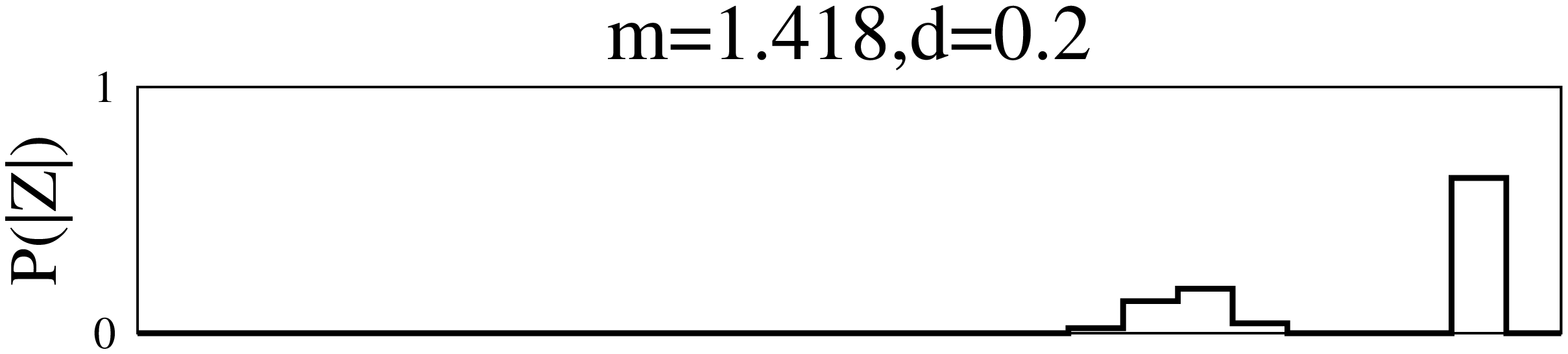}}\\
        \hskip 0.4 cm\resizebox{8cm}{!}{\includegraphics[angle=-90]{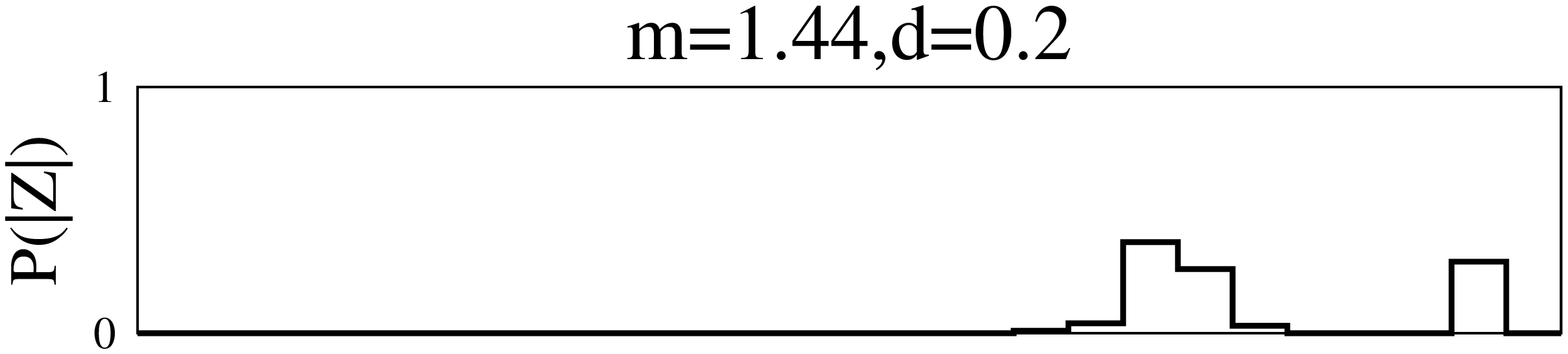}}\\
        \hskip 0.4 cm\resizebox{8cm}{!}{\includegraphics[angle=0]{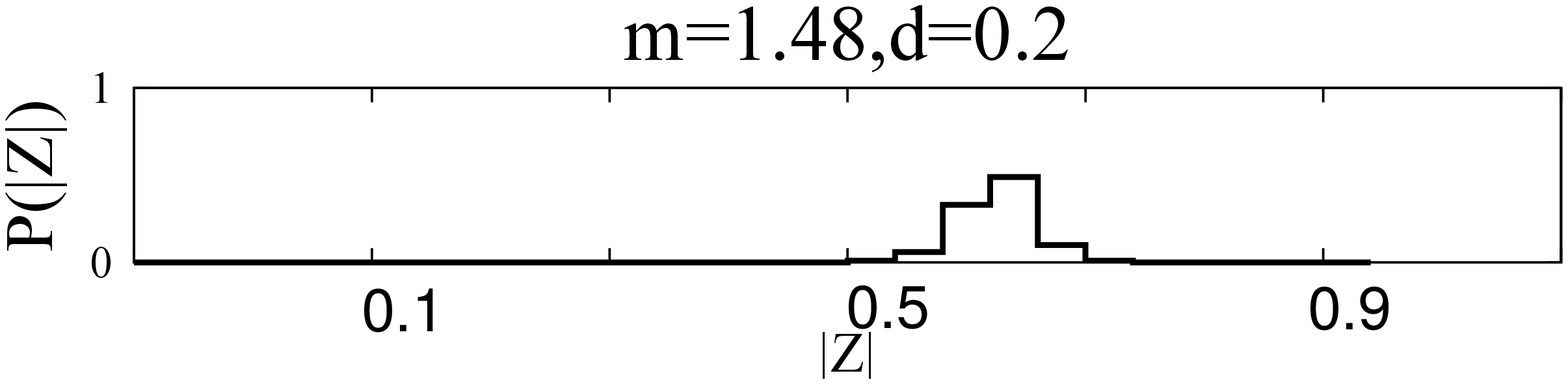}}\\

         \textbf{(b)} \resizebox{8cm}{!}{\includegraphics[angle=-90]{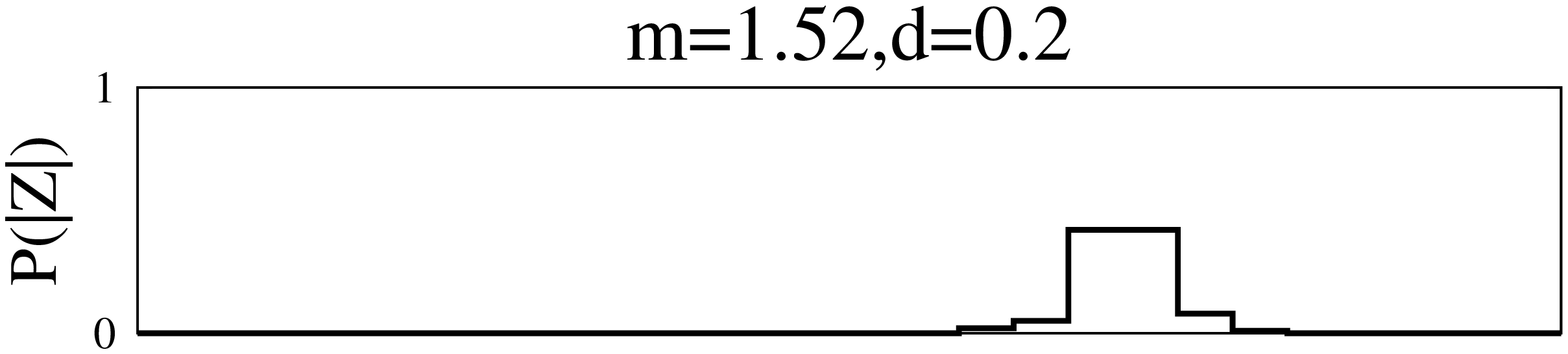}}\\
         \hskip 0.5 cm\resizebox{8cm}{!}{\includegraphics[angle=-90]{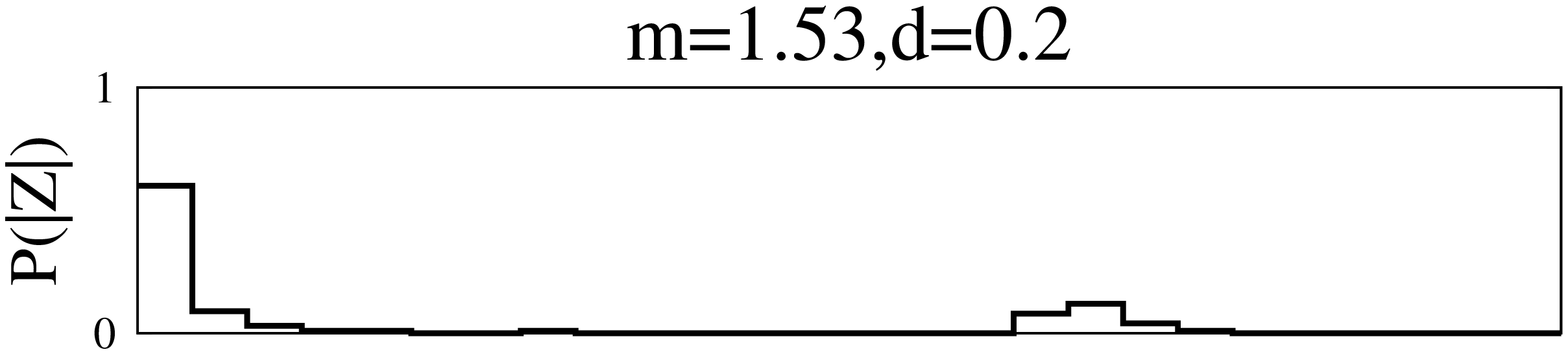}}\\
          \hskip 0.5 cm\resizebox{8cm}{!}{\includegraphics[angle=-90]{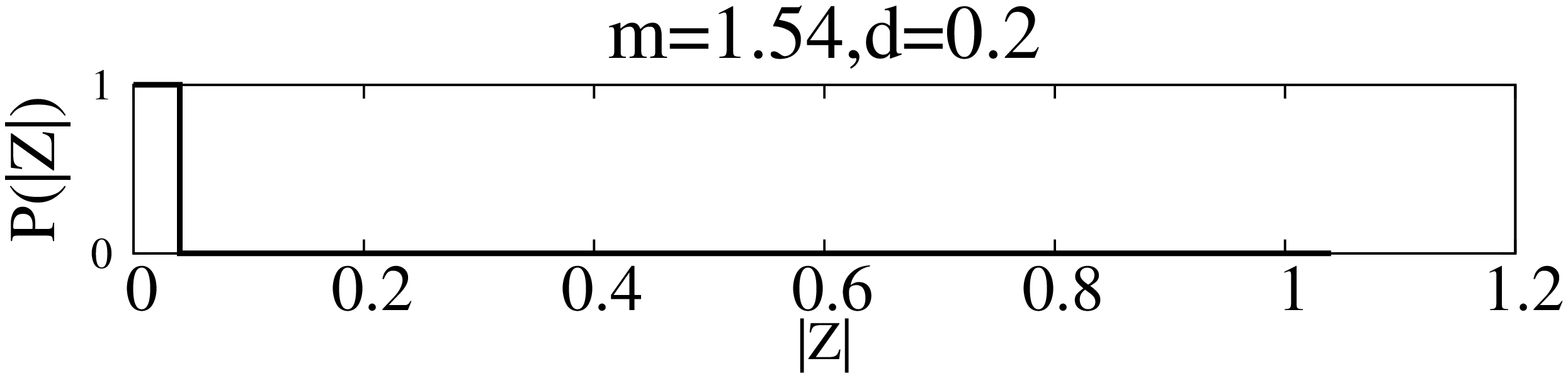}}\\
          \end{tabular}
\end{center}
\end{figure}

\begin{figure}[h]
\begin{center}
\begin{tabular}{c}
          \textbf{(c)}\resizebox{8cm}{!}{\includegraphics[angle=-90]{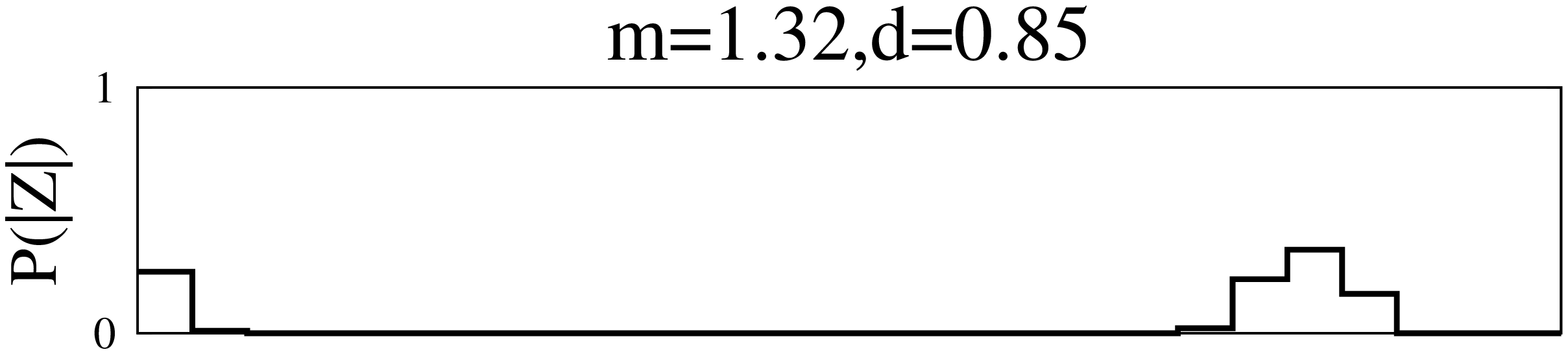}}\\
          \hskip 0.5 cm\resizebox{8cm}{!}{\includegraphics[angle=-90]{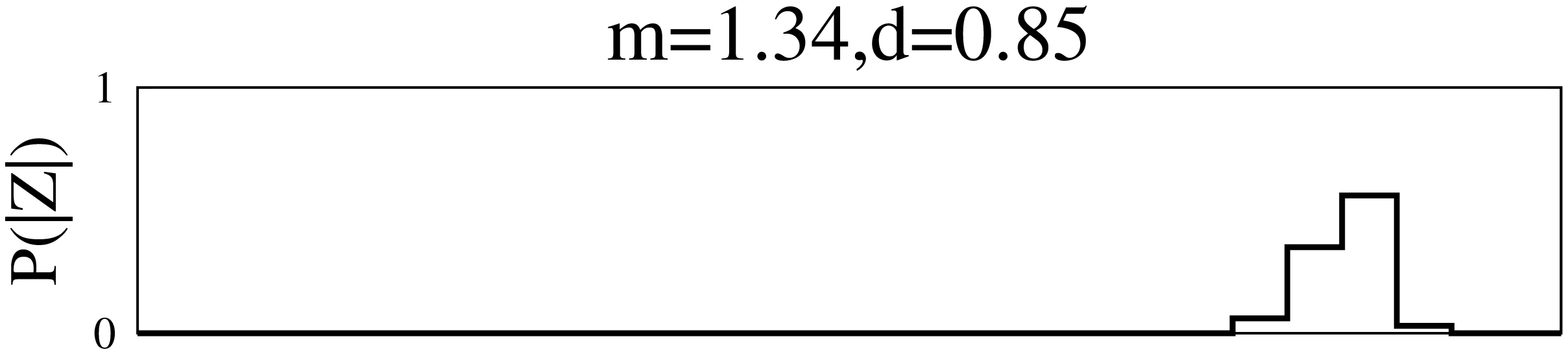}}\\
          \hskip 0.5 cm\resizebox{8cm}{!}{\includegraphics[angle=-90]{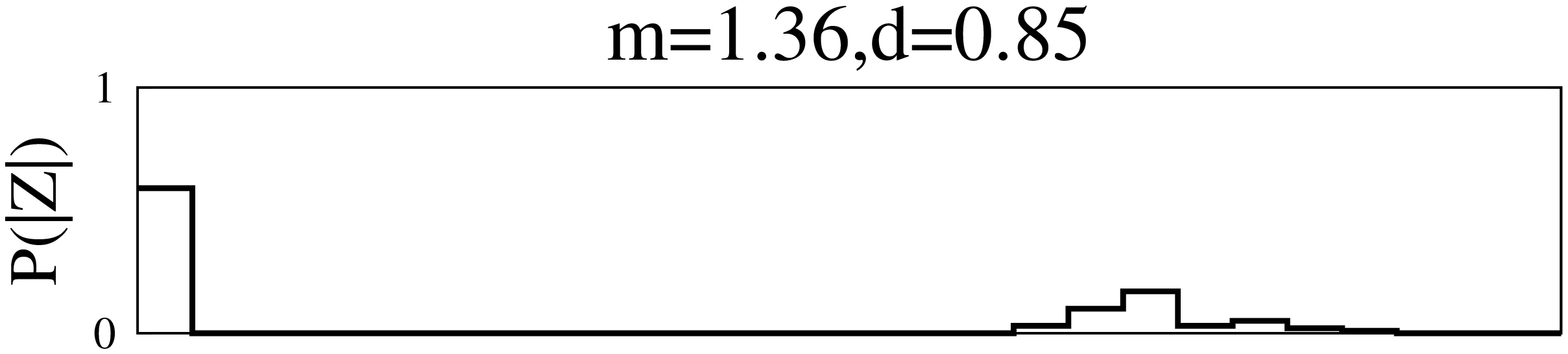}}\\
          \hskip 0.5 cm\resizebox{8cm}{!}{\includegraphics[angle=-90]{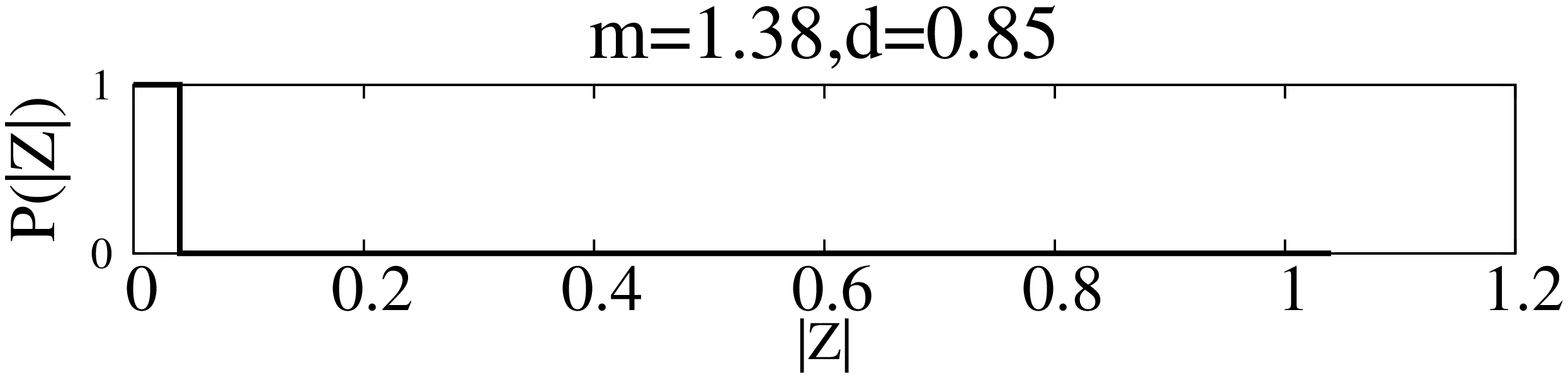}}\\
        
                  \end{tabular}
 \caption{{(Color Online) Histogram for the modulus of the order parameter $|Z|$ in quenched disorder at time 1000 unit with $N=100, r=0.35$ for an ensemble of 100 random initial phase distributions: (a) $d=0.2$  corresponding to a continuous transition, (b) $d=0.2$ corresponding to an abrupt synchronization and (c) abrupt synchronization for $d=0.85$.}}
\end{center}
\end{figure}

\newpage

\begin{figure}[h]
\begin{center}
\begin{tabular}{c}
        \textbf{(a)} \resizebox{8cm}{!}{\includegraphics[angle=-90]{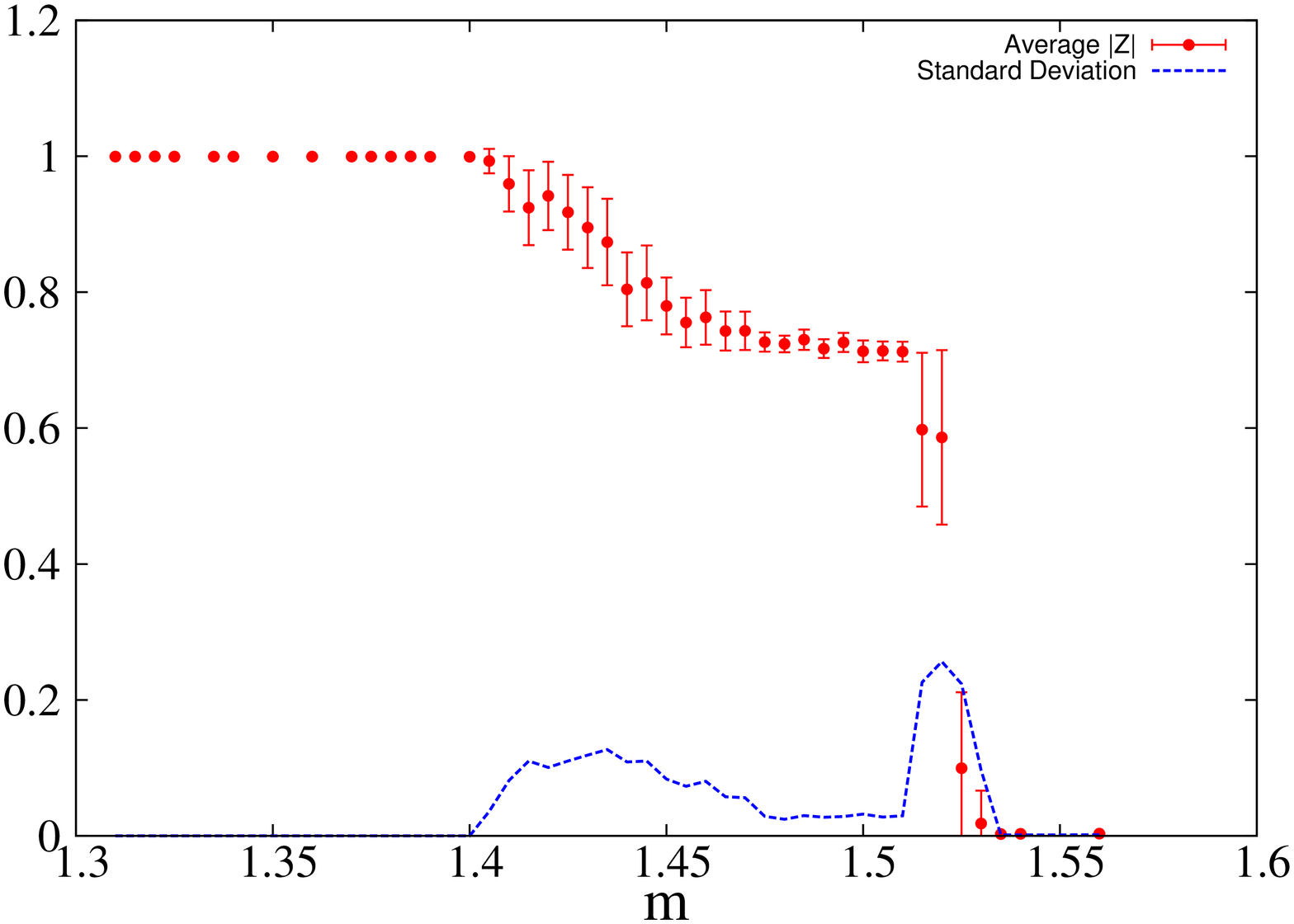}}
         \textbf{(b)}\resizebox{8cm}{!}{\includegraphics[angle=-90]{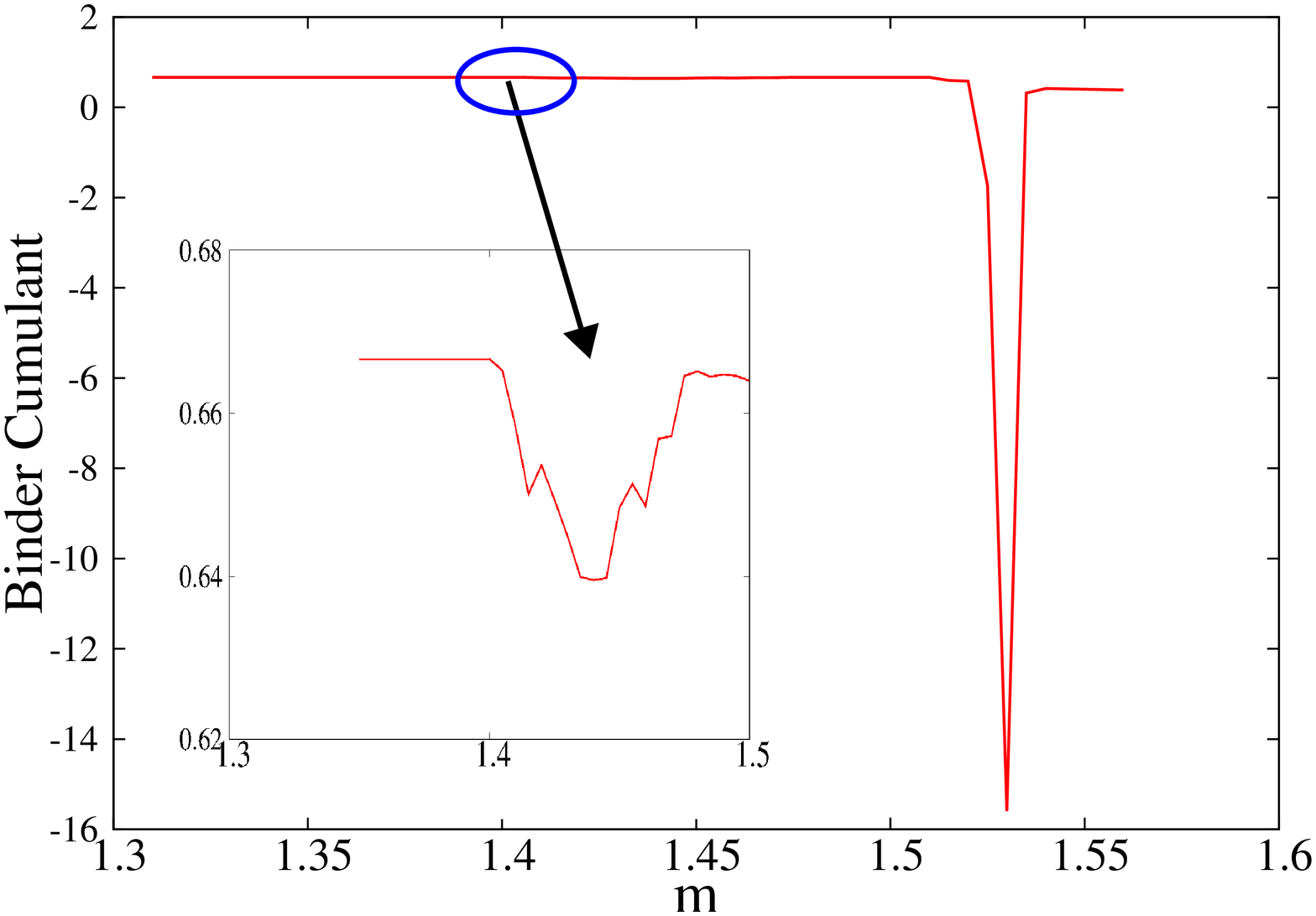}}\\
         \textbf{(a$^{\prime }$)} \resizebox{8cm}{!}{\includegraphics[angle=-90]{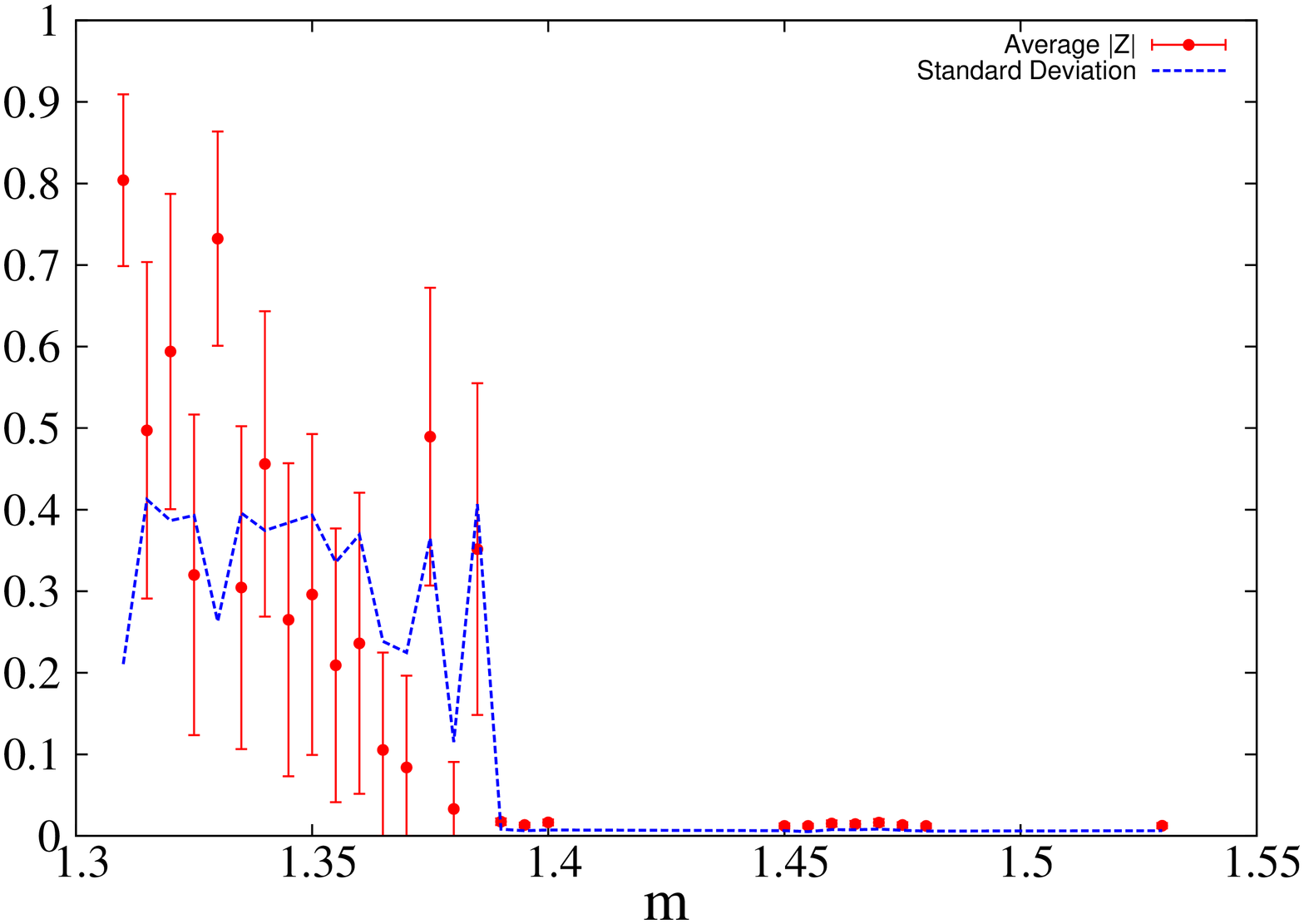}}
         \textbf{(b$^{\prime }$)}\resizebox{8cm}{!}{\includegraphics[angle=-90]{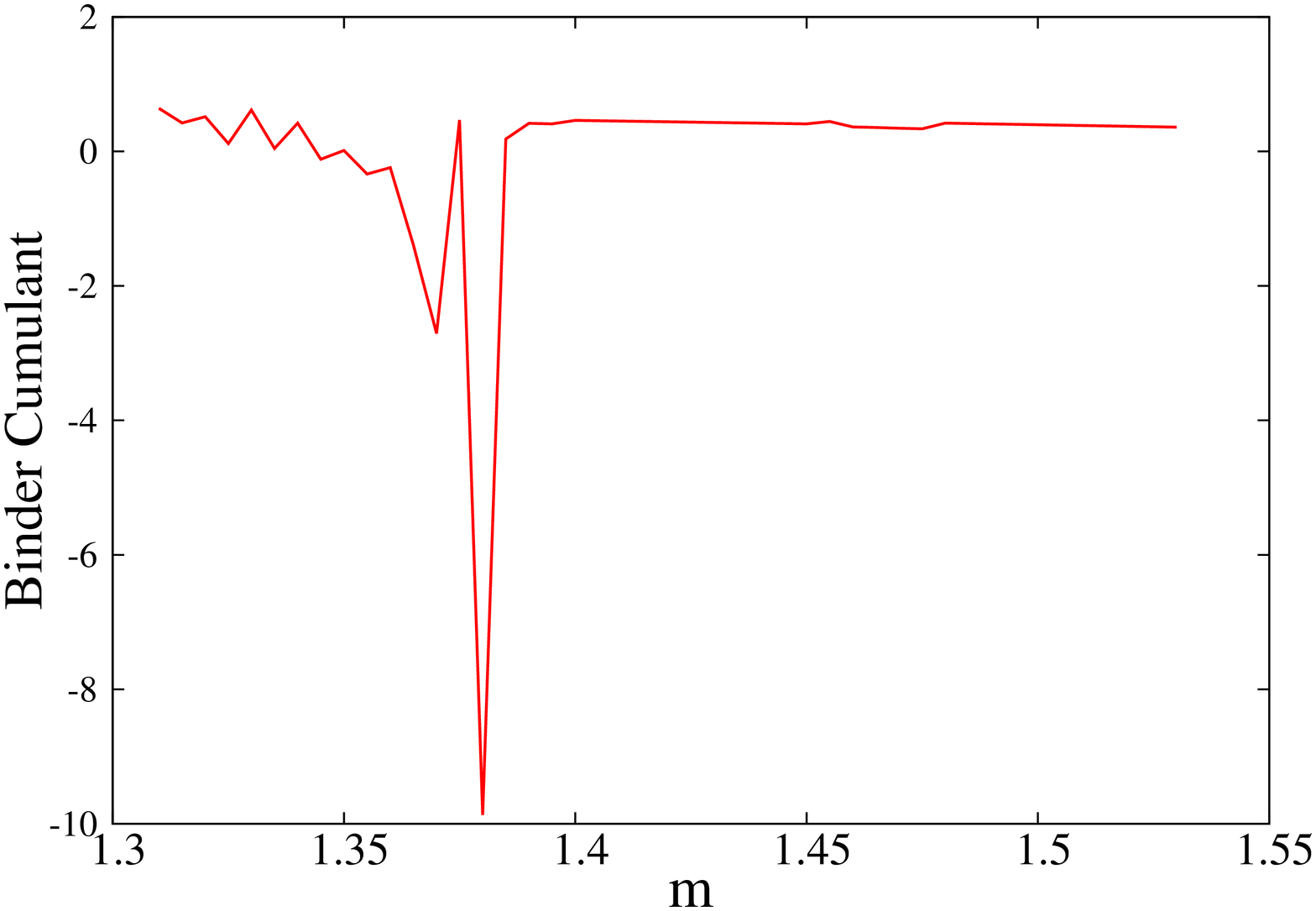}}
        
                  \end{tabular}
 \caption{{(Color Online) (a) Mean (with standard deviation as error bar) and Standard Deviation of the order parameter $|Z|$ and (b) Binder's fourth order cumulant for $N=100,r=0.35$, Gaussian quenched disorder with $d=0.2$ with $|Z|$ calculated after time 1000 unit and averaged over 100 random initial phase configurations but with identical values of the disordered Sakaguchi phases. The primed figures are same with $d=0.85$.}}
\end{center}
\end{figure}

\newpage

\begin{figure}[h]
\begin{center}
\begin{tabular}{c}
        \textbf{(a)} \resizebox{8cm}{!}{\includegraphics[angle=-90]{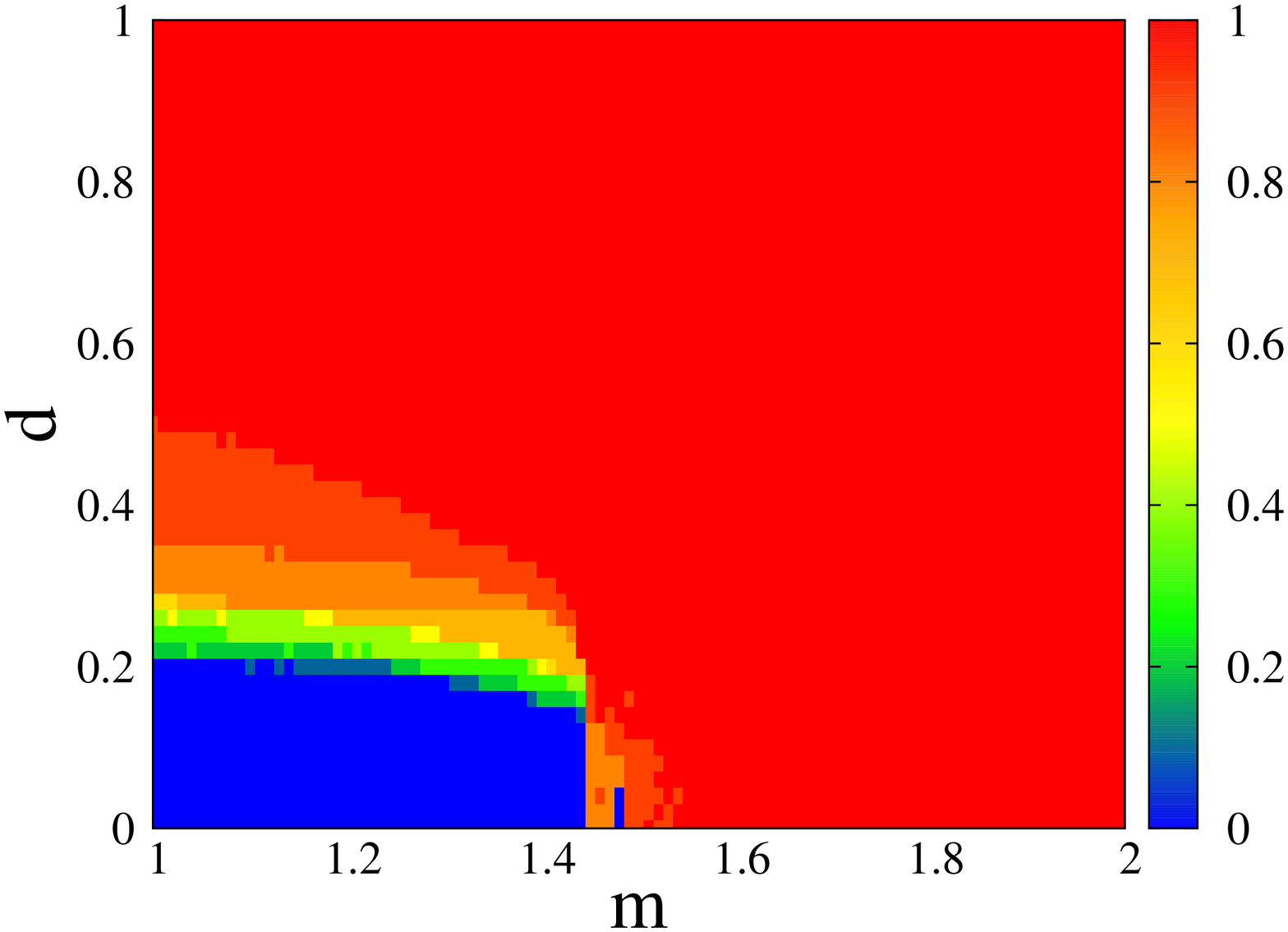}}
         \textbf{(a$^{\prime }$)}\resizebox{8cm}{!}{\includegraphics[angle=-90]{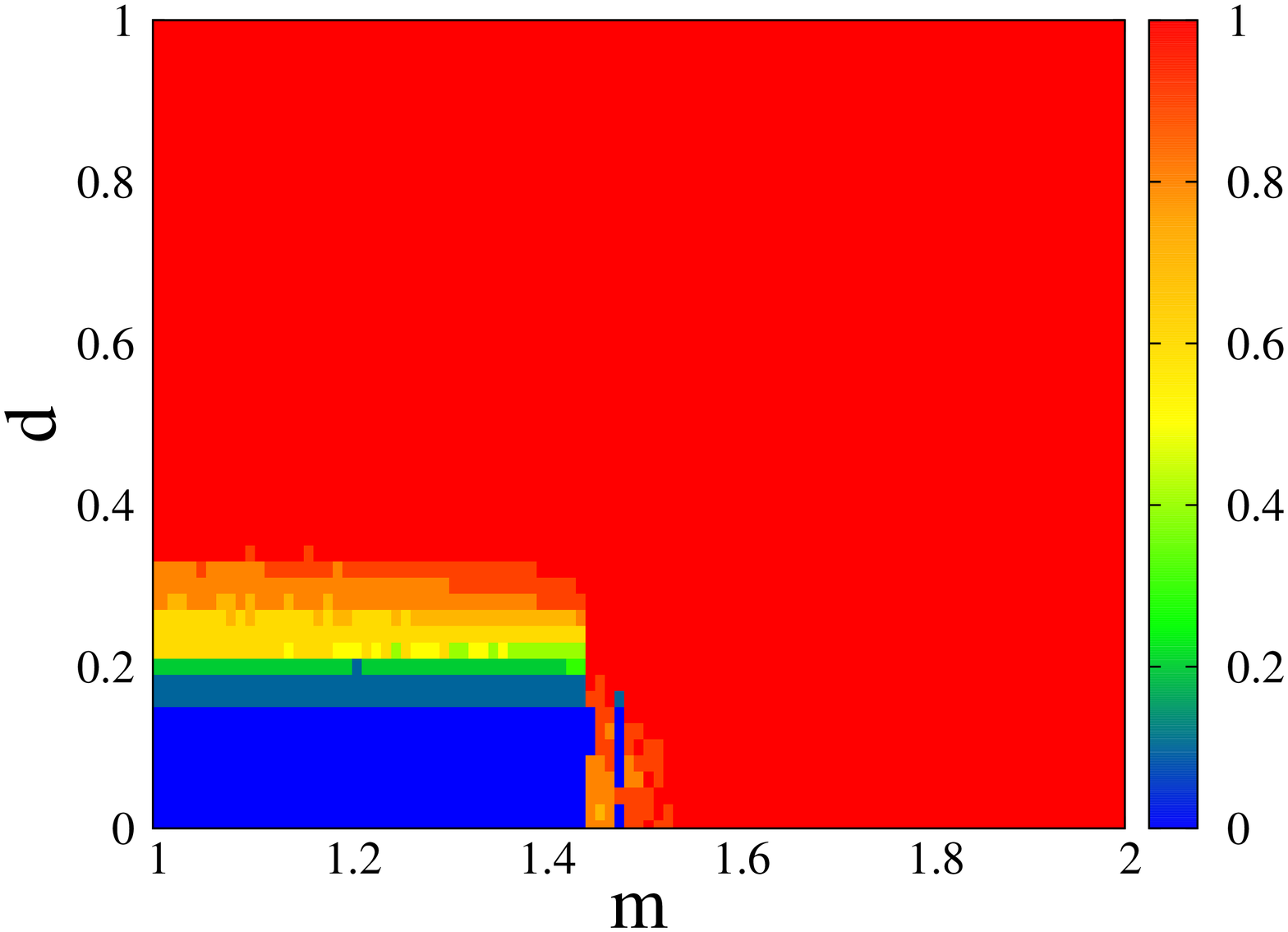}}\\
         \textbf{(b)} \resizebox{8cm}{!}{\includegraphics[angle=-90]{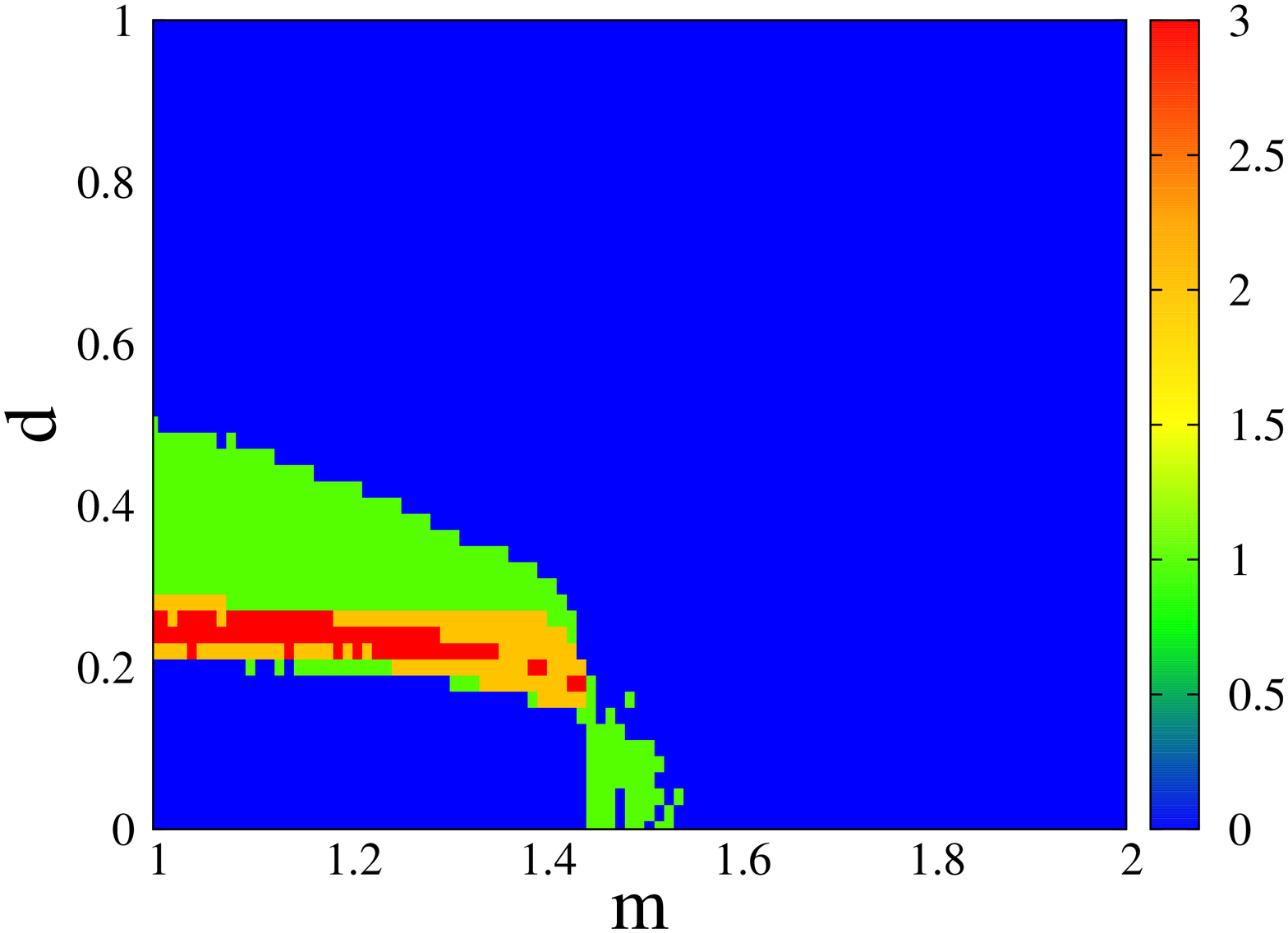}}
         \textbf{(b$^{\prime }$)}\resizebox{8cm}{!}{\includegraphics[angle=-90]{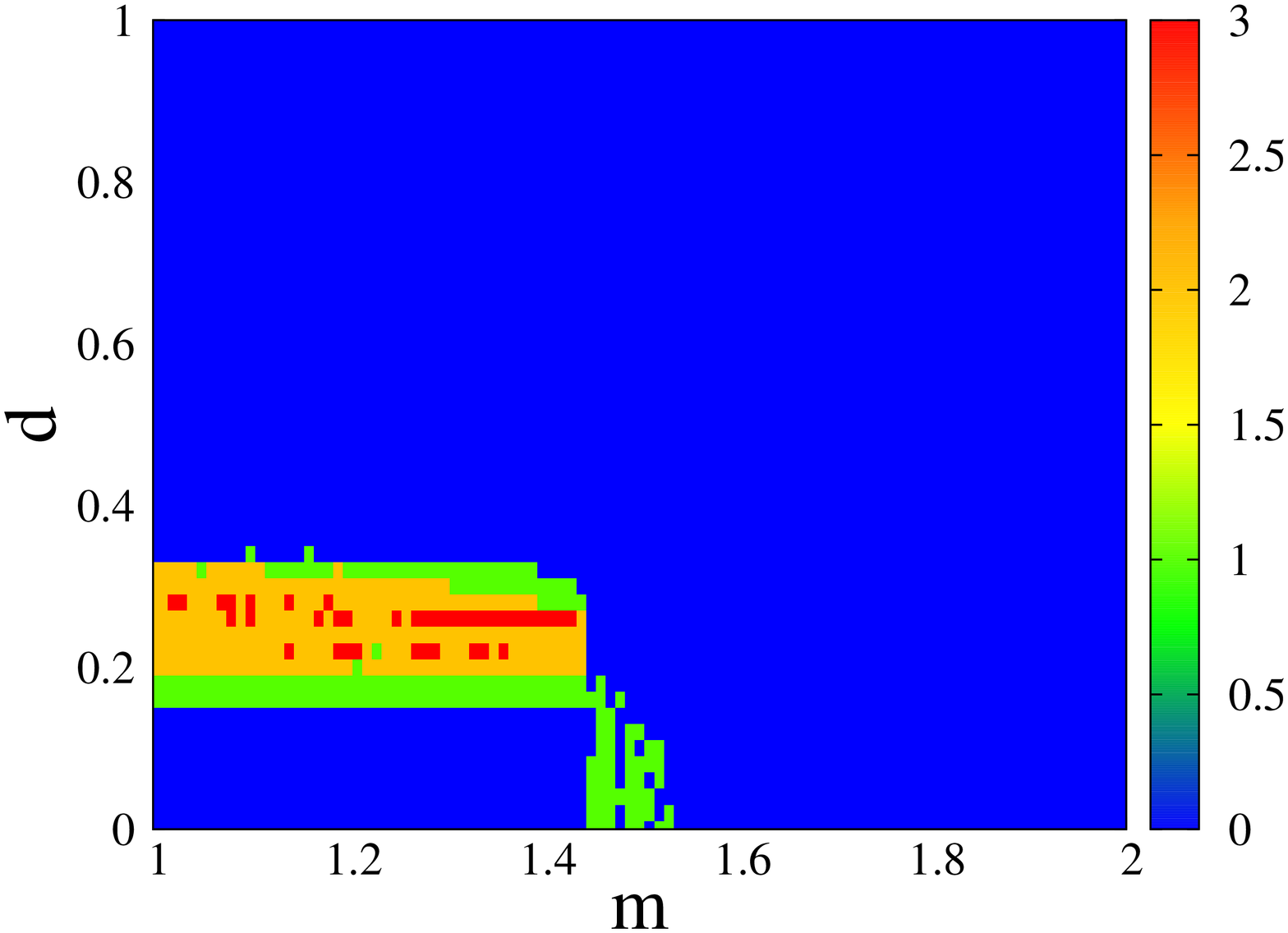}}
        
                  \end{tabular}
 \caption{{(Color Online) Phase Diagram for (a) Strength of Incoherence S and (b) Discontinuity Measure $\eta$ with time-averaging performed between times 800 to 1000 units for Gaussian quenched disorder, with $N=100, r=0.35, N_{g}=10, \delta =0.1$  with identical initial random configuration of phases showing a band around $m=1.5$ for lower values of $d$ corresponding to chimera states; the plots with primed labels are the same but with Discrete Bimodal quenched disorder.}}
\end{center}
\end{figure}
\newpage
\begin{figure}[h]
\begin{center}
\begin{tabular}{c}
        \resizebox{8cm}{!}{\includegraphics[angle=-0]{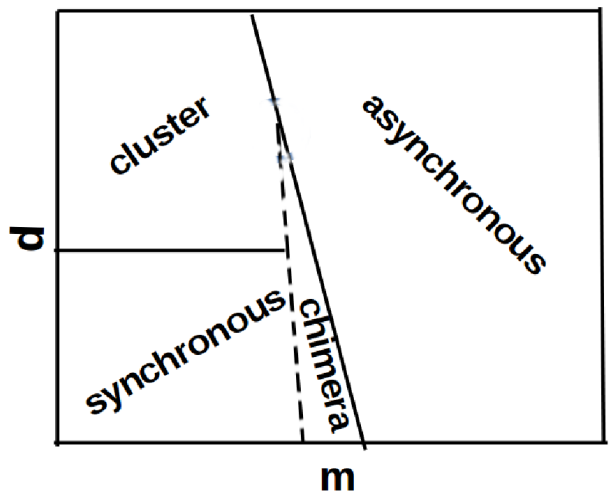}}

                  \end{tabular}
 \caption{{Schematic Phase Diagram with non-local coupling and quenched disorder, the dashed boundary corresponds to continuous transition.}}
\end{center}
\end{figure}
\newpage
\begin{figure}[h]
\begin{center}
\begin{tabular}{c}
        \textbf{(a)} \resizebox{8cm}{!}{\includegraphics[angle=-90]{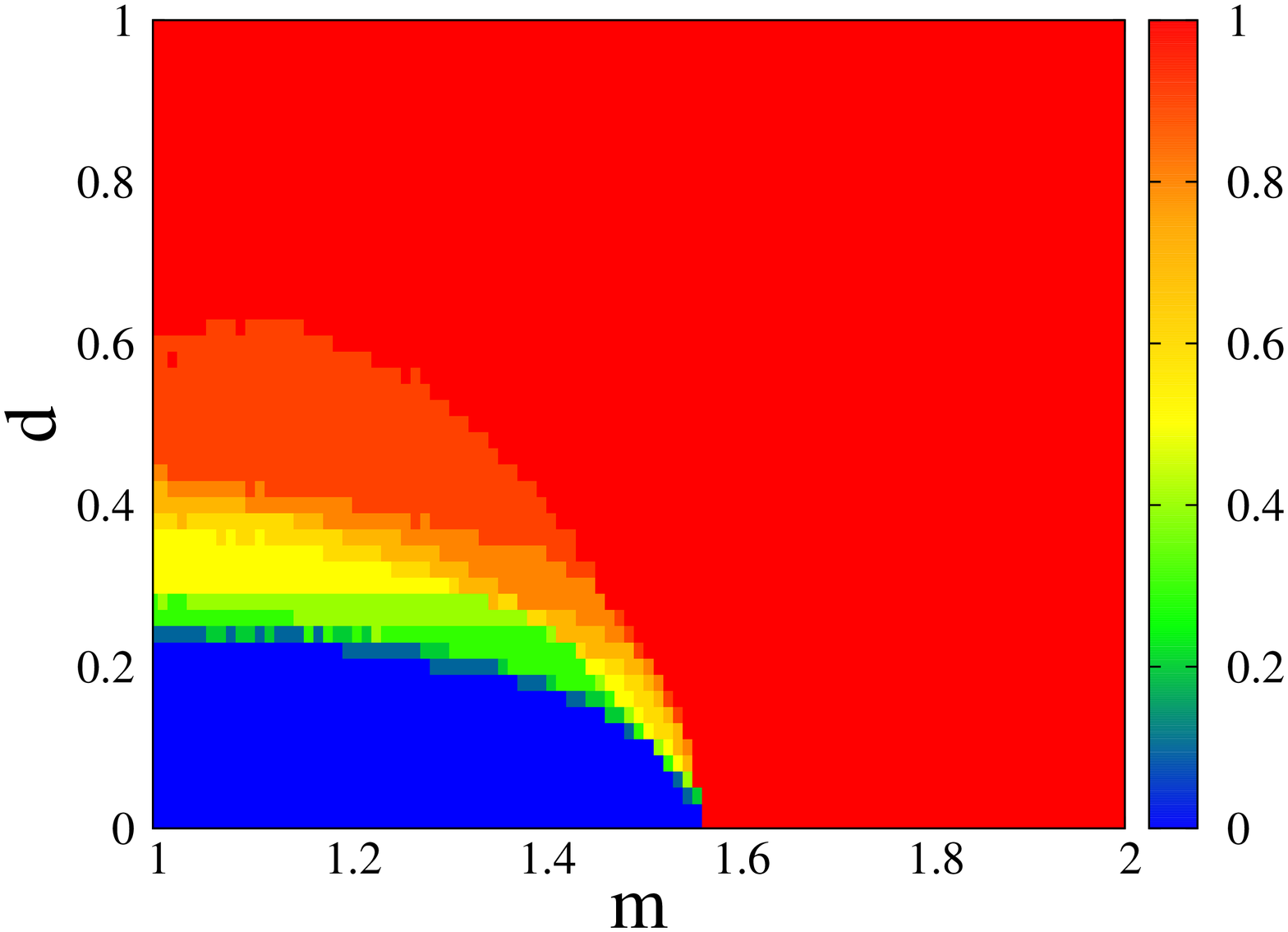}}
         \textbf{(a$^{\prime }$)}\resizebox{8cm}{!}{\includegraphics[angle=-90]{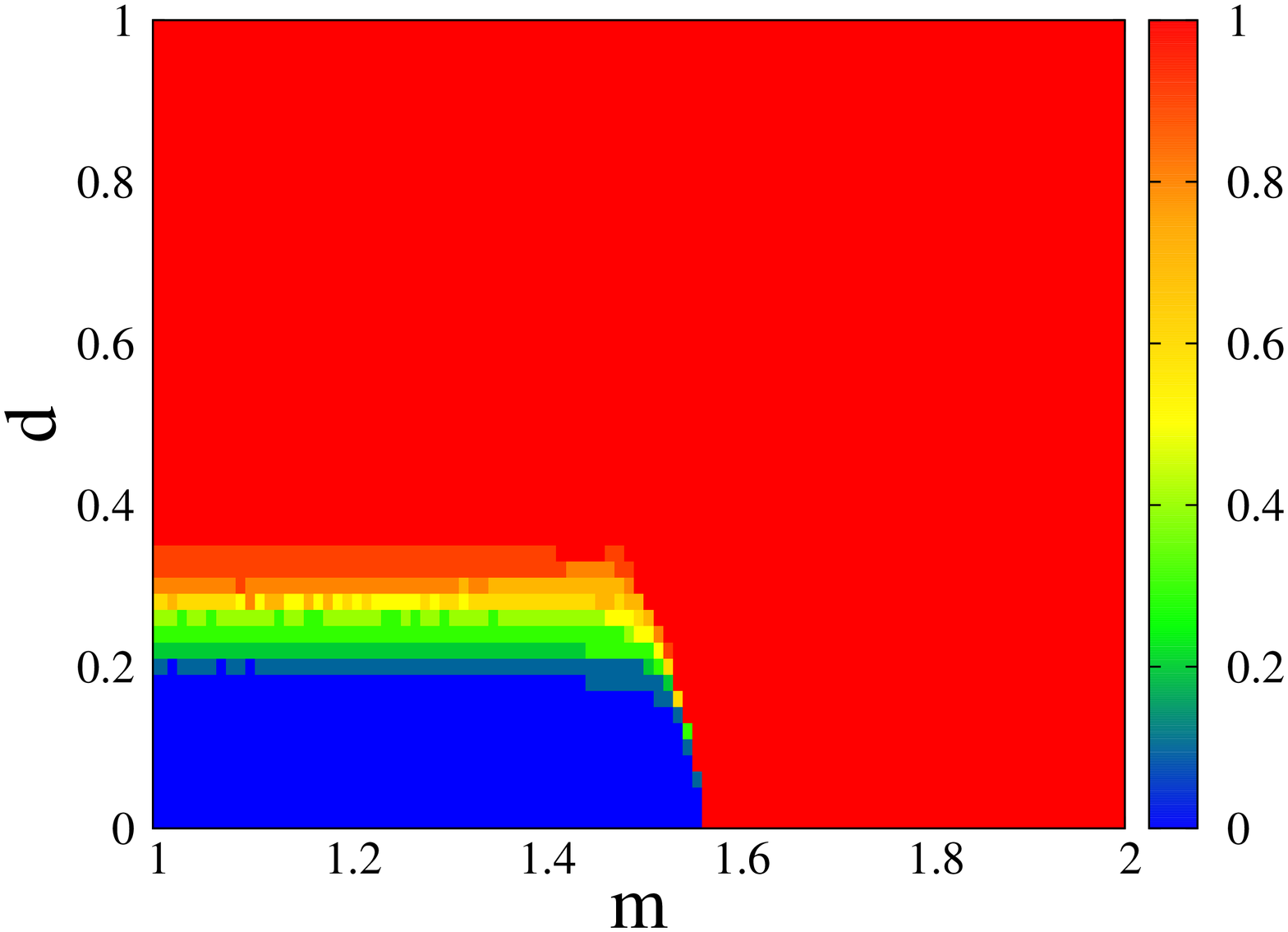}}\\
         \textbf{(b)} \resizebox{8cm}{!}{\includegraphics[angle=-90]{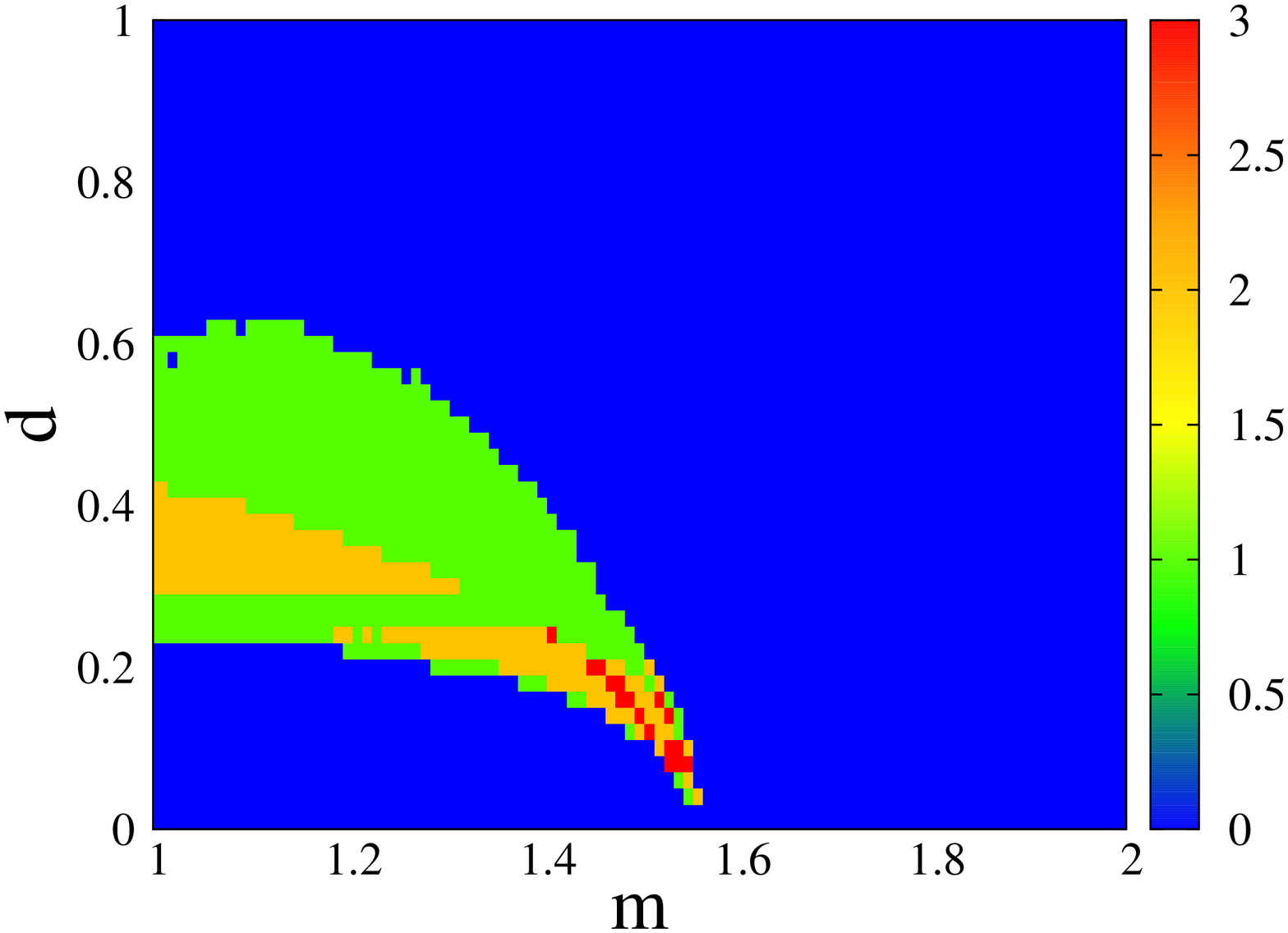}}
         \textbf{(b$^{\prime }$)}\resizebox{8cm}{!}{\includegraphics[angle=-90]{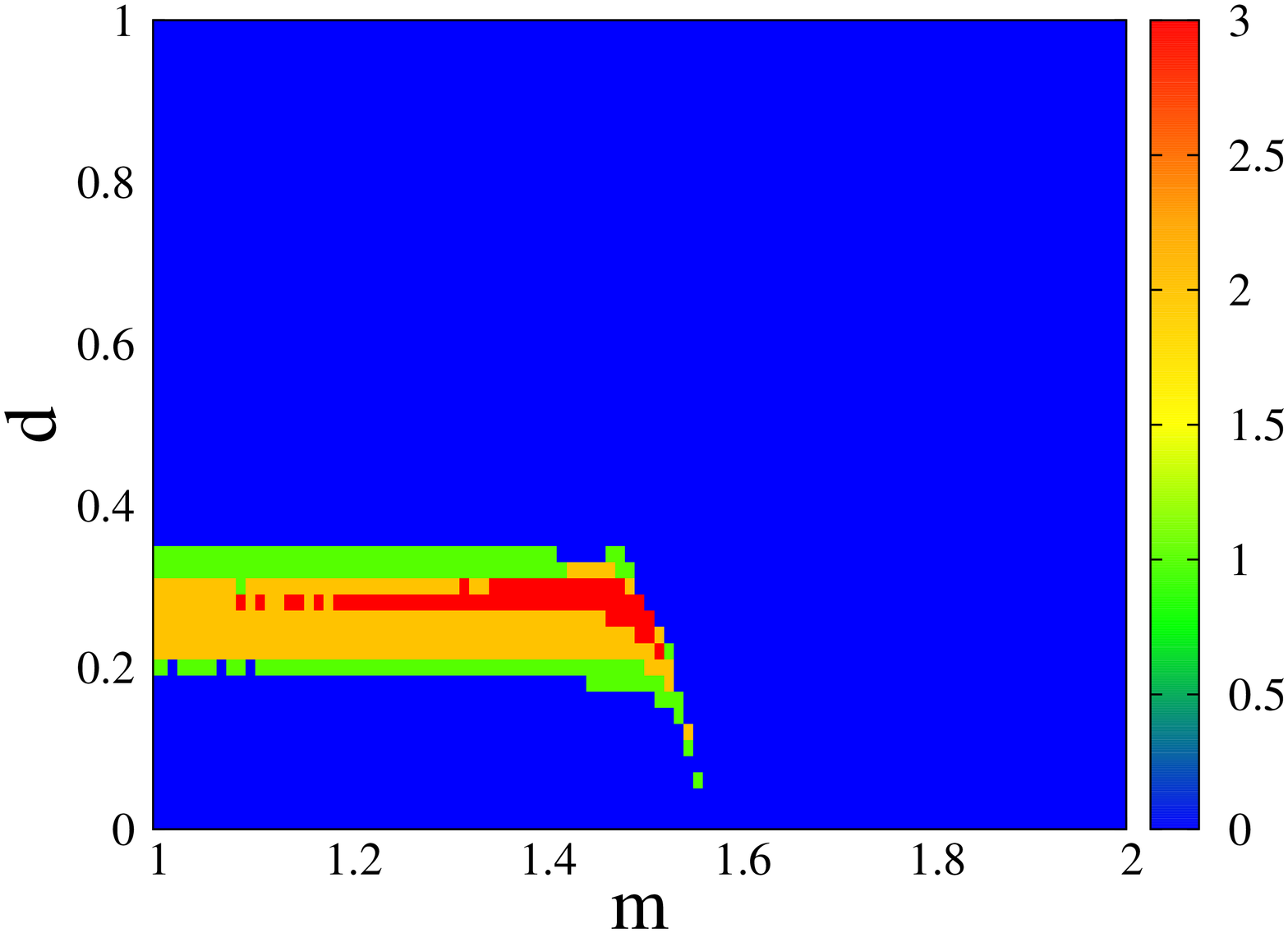}}
        
                  \end{tabular}
 \caption{{(Color Online) Phase Diagram for (a) Strength of Incoherence $S$ and (b) Discontinuity Measure $\eta$ with time-averaging performed between times 800 to 1000 units for Gaussian quenched disorder, with $N=100, r=0.49, N_{g}=10, \delta =0.1$  with identical initial random configuration of phases showing a tapering around $m=1.5$ for lower values of $d$ corresponding to a sharp and discontinuous transition; the plots with primed labels are the same but with Discrete Bimodal quenched disorder.}}
\end{center}
\end{figure}

\newpage

\begin{figure}[h]
\begin{center}
\begin{tabular}{c}
        \textbf{(a)} \resizebox{8cm}{!}{\includegraphics[angle=-90]{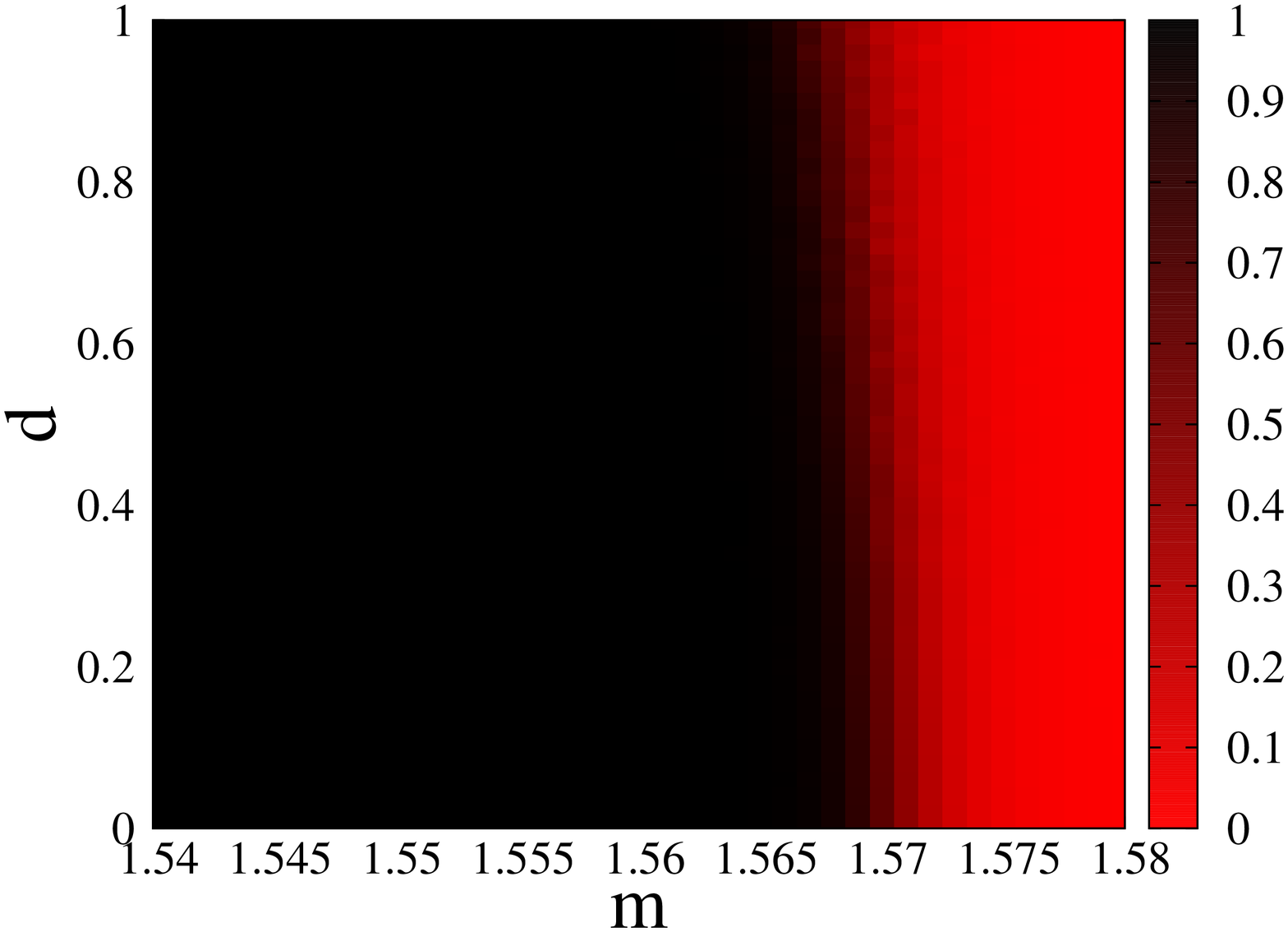}}
         \textbf{(b)}\resizebox{8cm}{!}{\includegraphics[angle=-90]{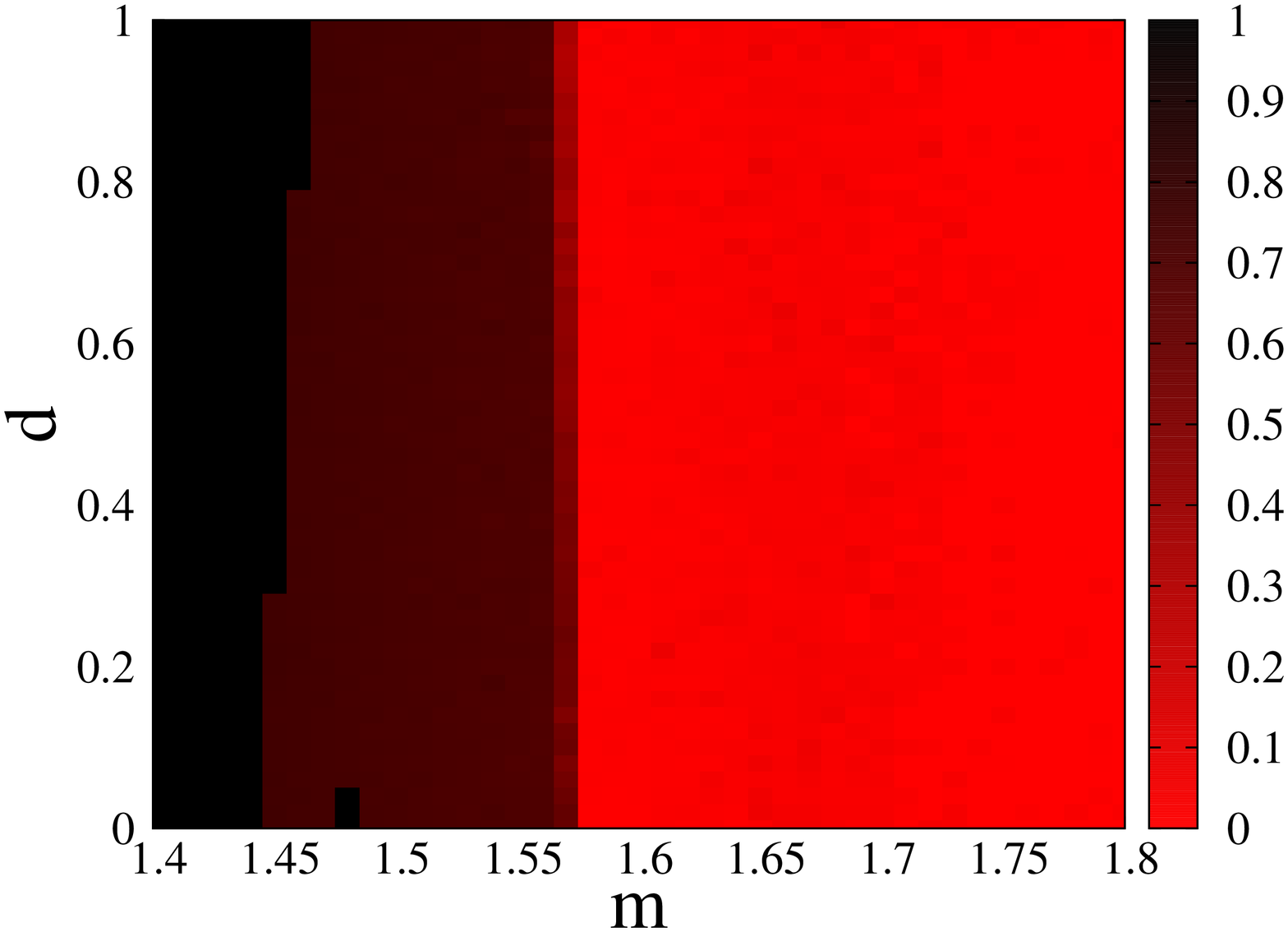}}
        
                  \end{tabular}
 \caption{{(Color Online) Phase Diagram for modulus of the order parameter $|Z|$ at time 1000 unit for (a) $r=0.49$ and (b) $r=0.35$, with $N=100$ with annealed disorder.}}
\end{center}
\end{figure}
\newpage
\begin{figure}[h]
\begin{center}
\begin{tabular}{c}
        \textbf{(a)} \resizebox{8cm}{!}{\includegraphics[angle=-90]{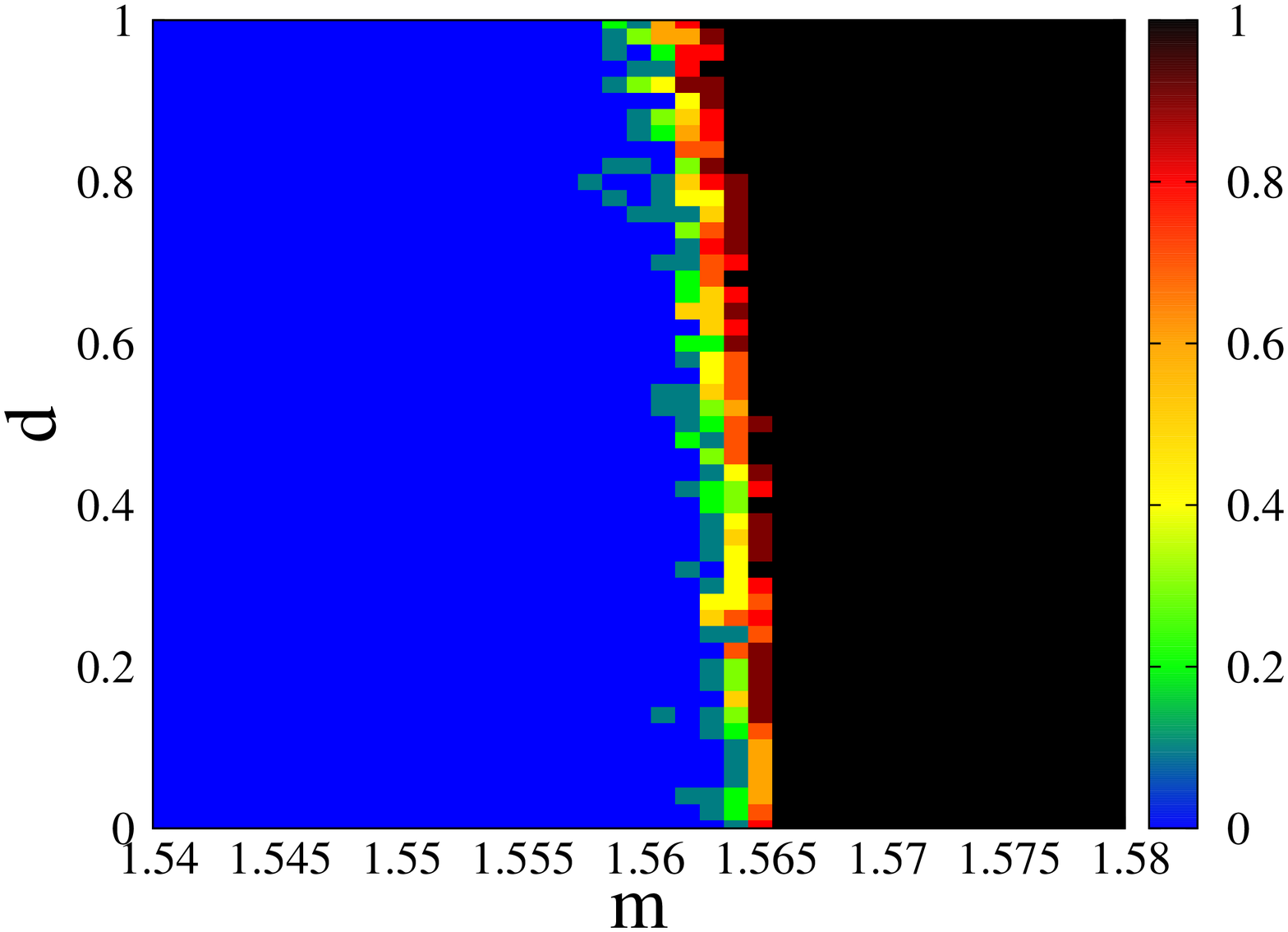}}
         \textbf{(a$^{\prime }$)}\resizebox{8cm}{!}{\includegraphics[angle=-90]{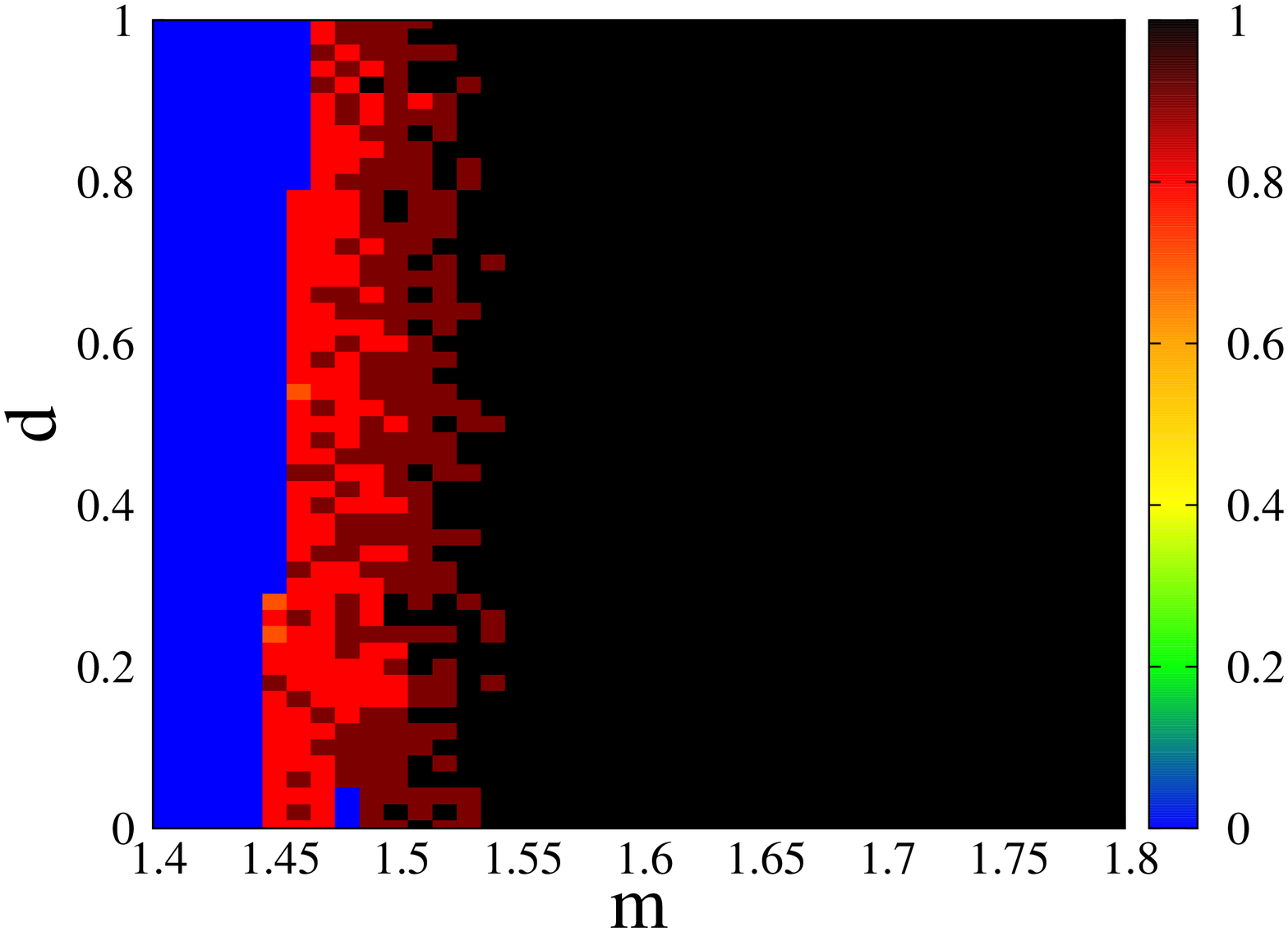}}\\
         \textbf{(b)} \resizebox{8cm}{!}{\includegraphics[angle=-90]{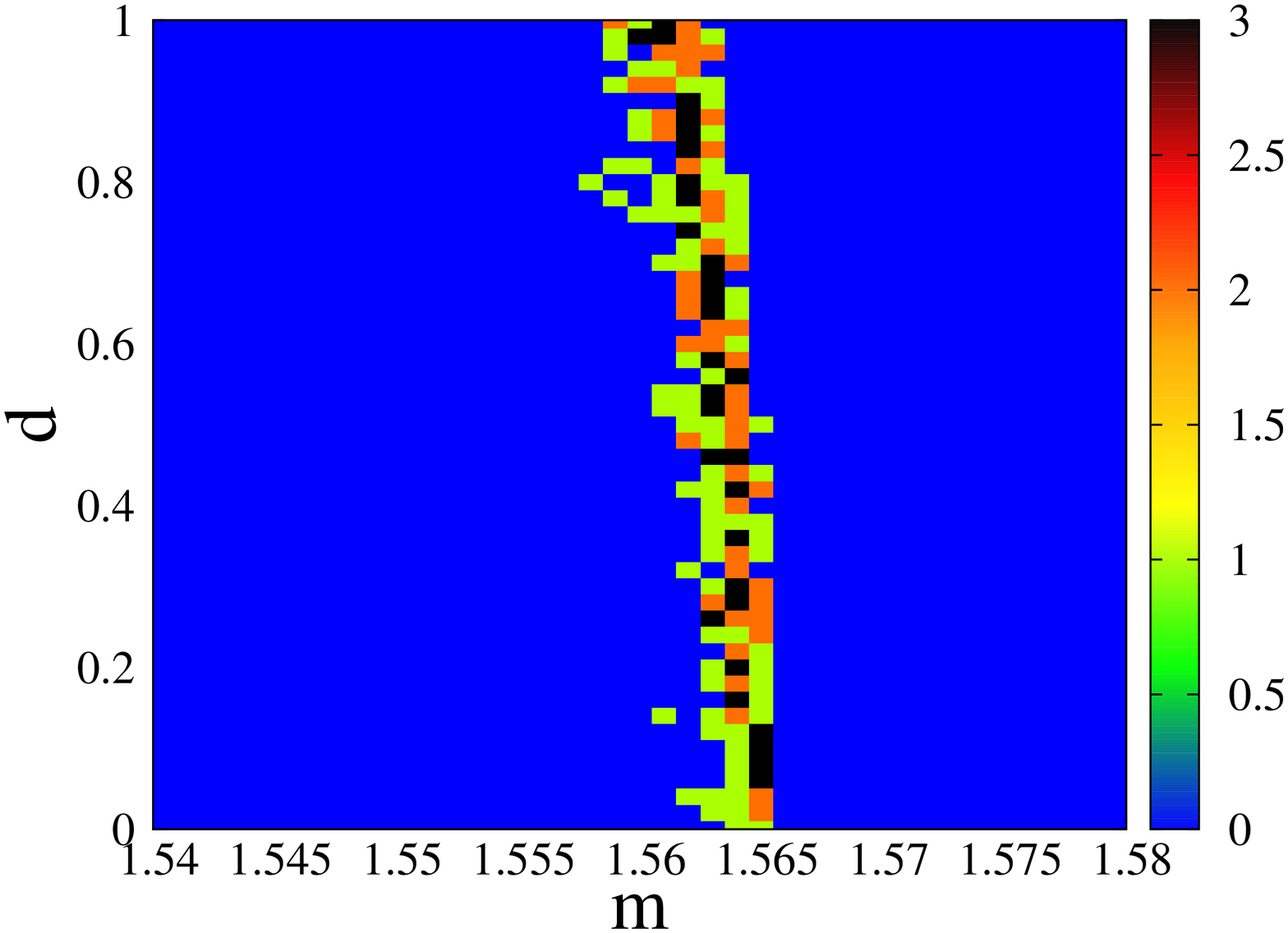}}
         \textbf{(b$^{\prime }$)}\resizebox{8cm}{!}{\includegraphics[angle=-90]{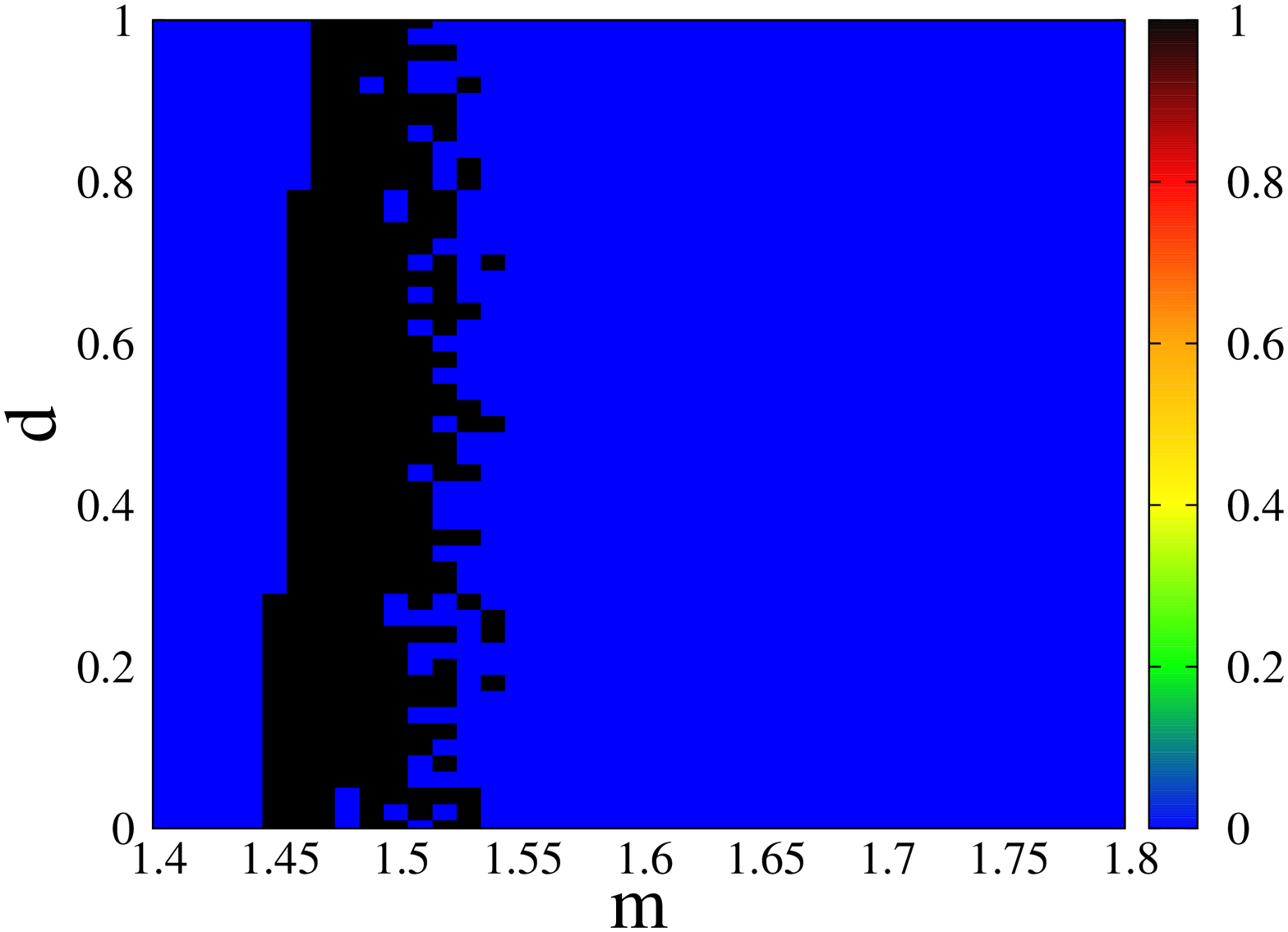}}
        
                  \end{tabular}
 \caption{{(Color Online) Phase Diagram for (a) Strength of Incoherence $S$ and (b) Discontinuity Measure $\eta$ with time-averaging performed between times 800 to 1000 units for annealed disorder, with $N=100, r=0.49, N_{g}=10, \delta =0.1$  with identical initial random configuration of phases; the plots with primed labels are the same but with non-local coupling where $r=0.35$.}}
\end{center}
\end{figure}
\newpage
\begin{figure}[h]
\begin{center}
\begin{tabular}{c}
        \textbf{(a)} \resizebox{3 cm}{!}{\includegraphics[angle=-90]{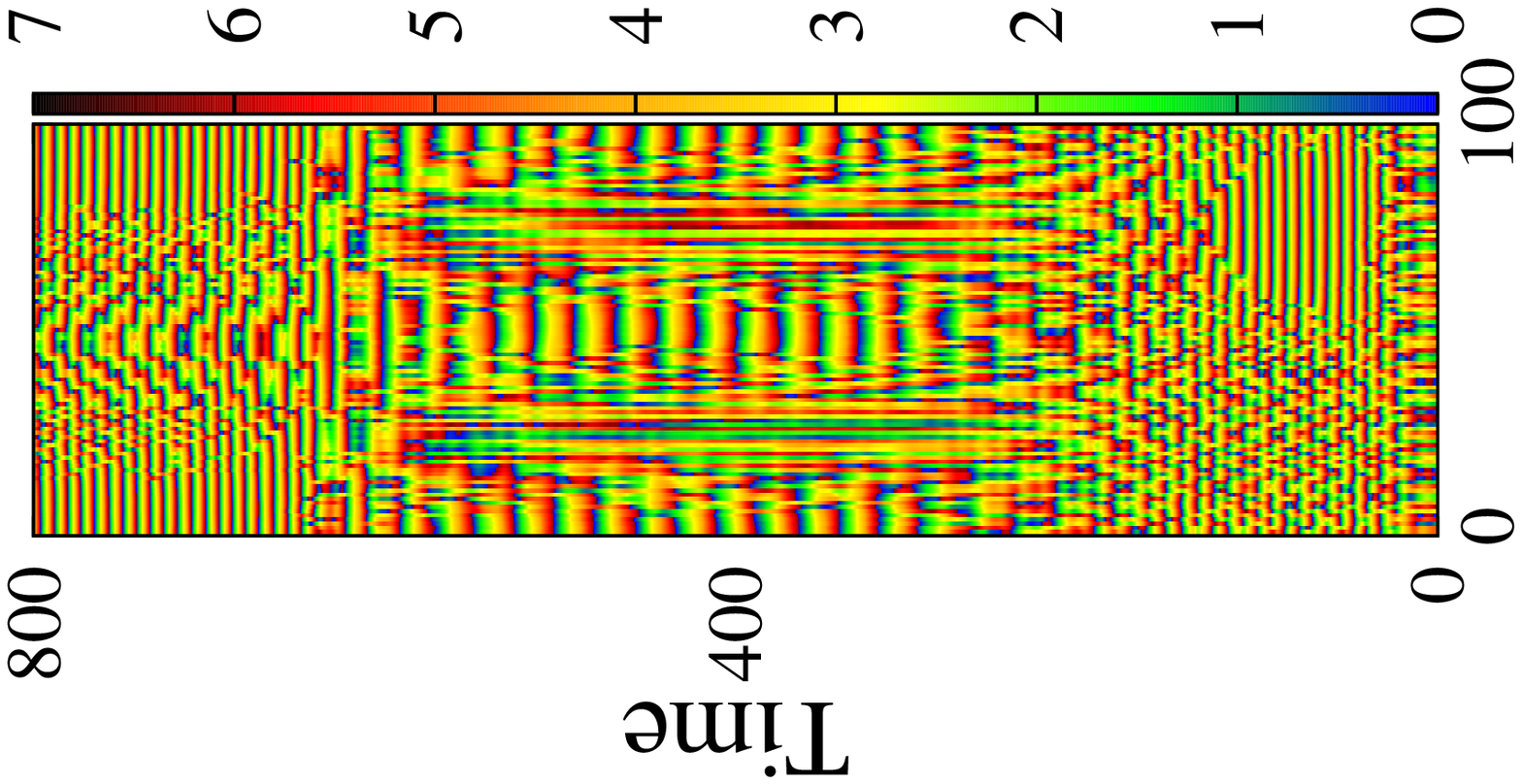}}
         \textbf{(b)}\resizebox{3 cm}{!}{\includegraphics[angle=-90]{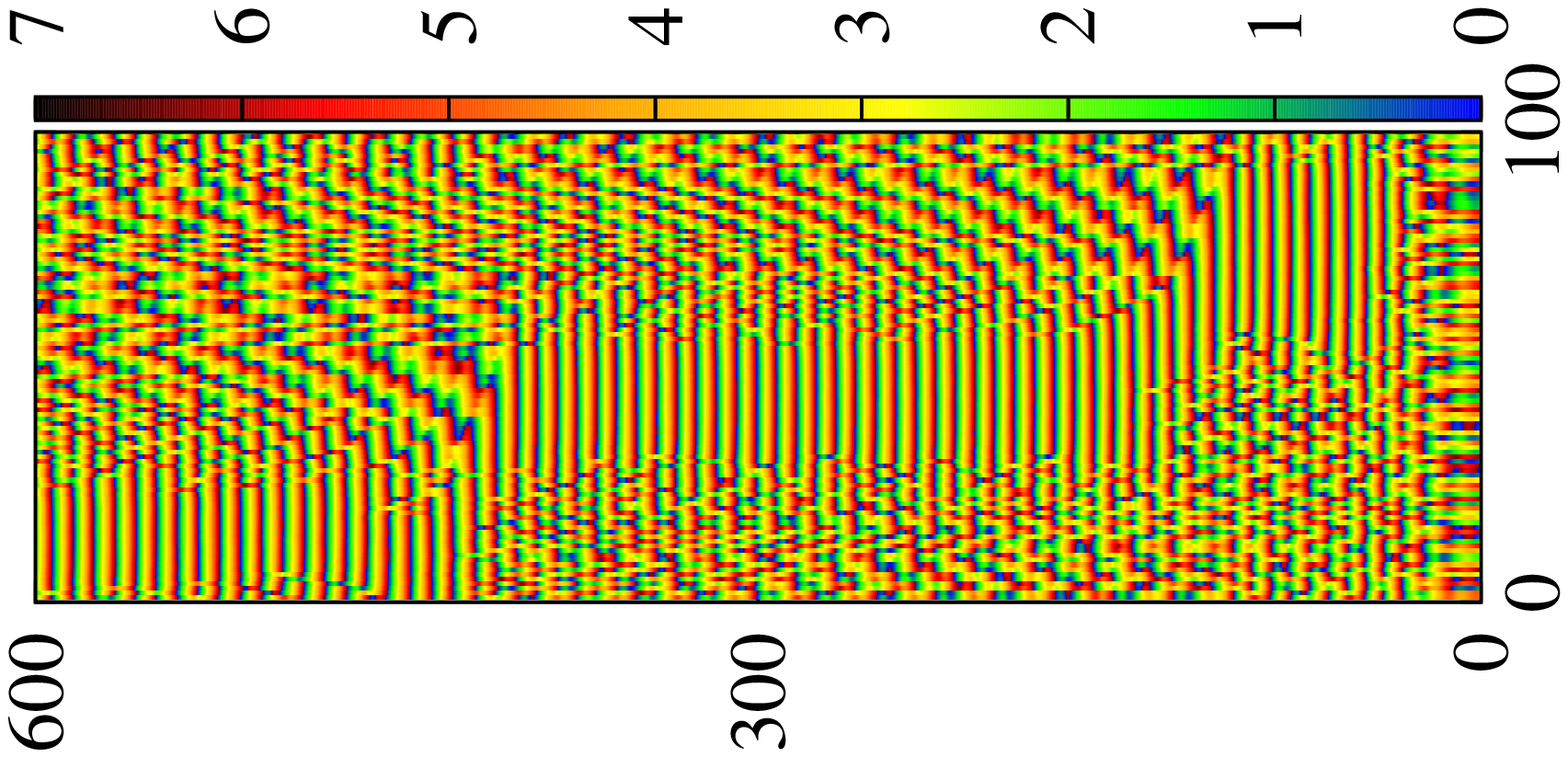}}
         \textbf{(c)}\resizebox{3 cm}{!}{\includegraphics[angle=-90]{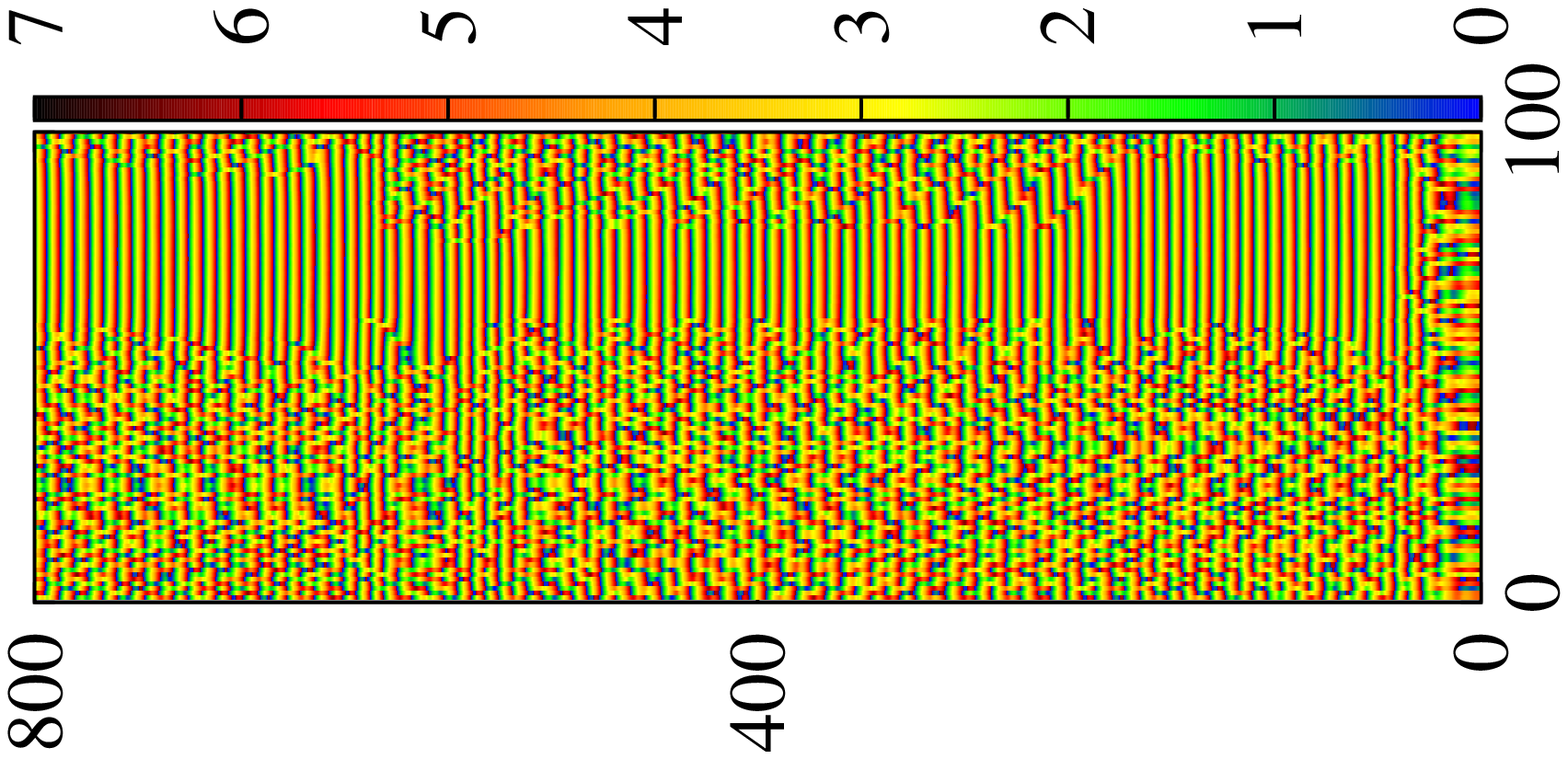}}\\
           \textbf{ Oscillator Index}
                          \end{tabular}
 \caption{{(Color Online) Dynamics of chimera states in space (x-axis) and time (y-axis) with $N=100$: (a) for $t=0-100$ and $501-800, \alpha=1.46$ for all oscillators; for other times, $\alpha=2.0$ for oscillator pairs present at lattice sites $10-40$ and $60-90$, while $\alpha=1.46$ for the rest (b) for $t=0-100$, $\alpha=1.46$ for all oscillators; for $t=101-400$, $\alpha=2.0$ for oscillator pairs present at lattice sites $1-20$ and $70-99$, while $\alpha=1.46$ for the rest; for $t=401-600$, $\alpha=2.0$ for oscillator pairs present at lattice sites $40-90$, while $\alpha=1.46$ for the rest (c) for $t=0-200$ and $601-800, \alpha=1.46$ for all oscillators; for other times, $\alpha=2.0$ for oscillator pairs present at lattice sites $40-60$ and $80-99$, while $\alpha=1.46$ for the rest.}}
\end{center}
\end{figure}
\newpage
\begin{figure}[h]
\begin{center}
\begin{tabular}{c}
        \textbf{(a)} \resizebox{3.1 cm}{!}{\includegraphics[angle=-90]{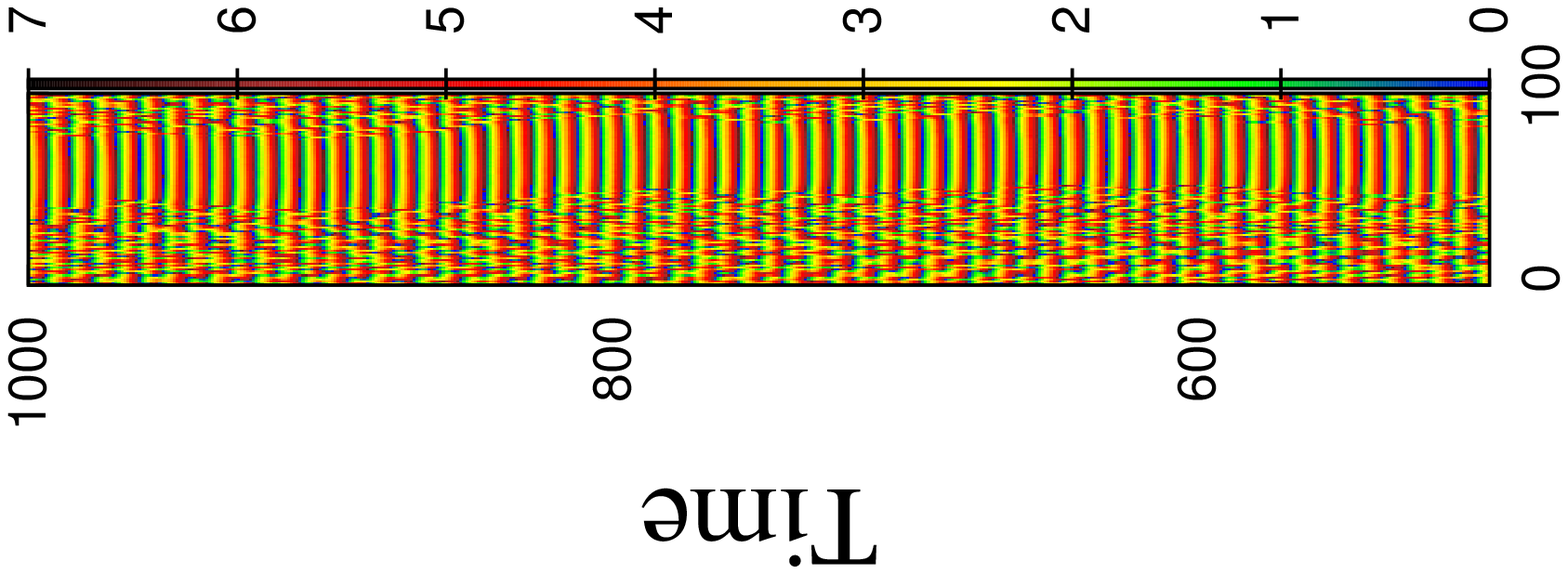}}
         \textbf{(b)}\resizebox{3 cm}{!}{\includegraphics[angle=-90]{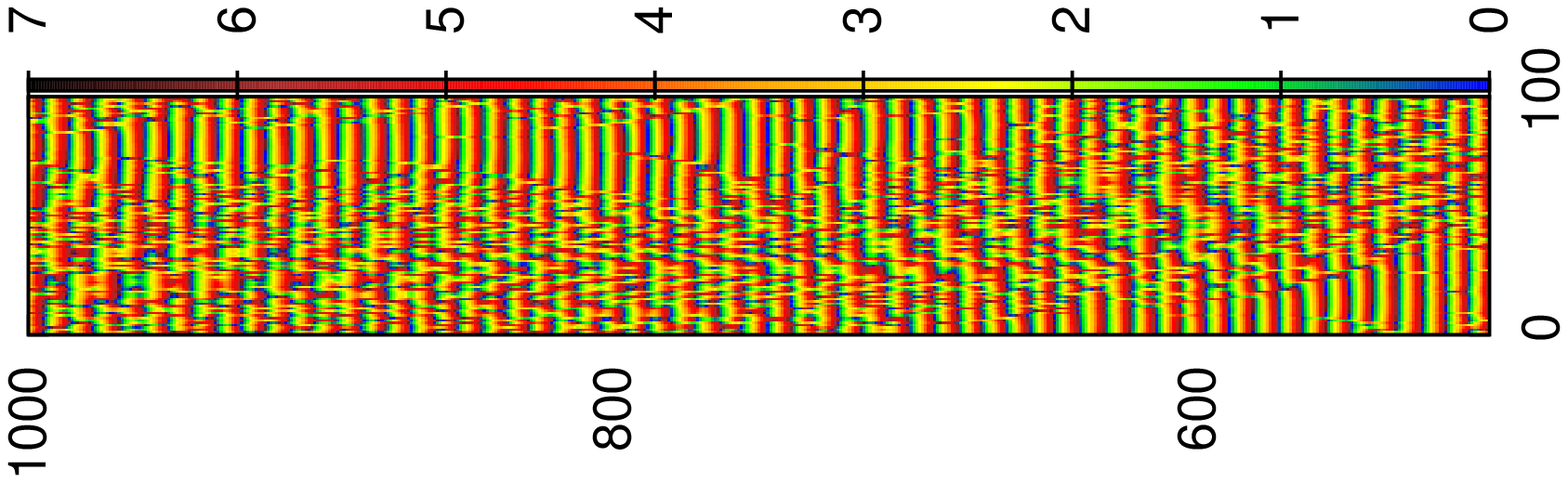}}
         \textbf{(c)}\resizebox{3 cm}{!}{\includegraphics[angle=-90]{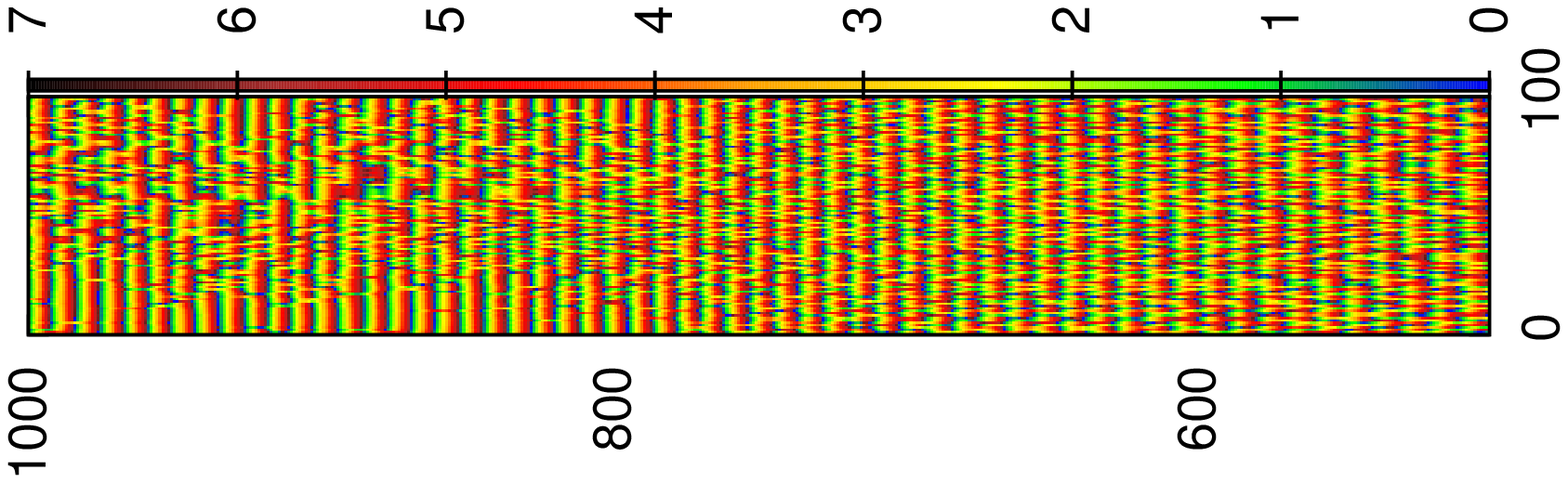}}
          \textbf{(d)}\resizebox{3 cm}{!}{\includegraphics[angle=-90]{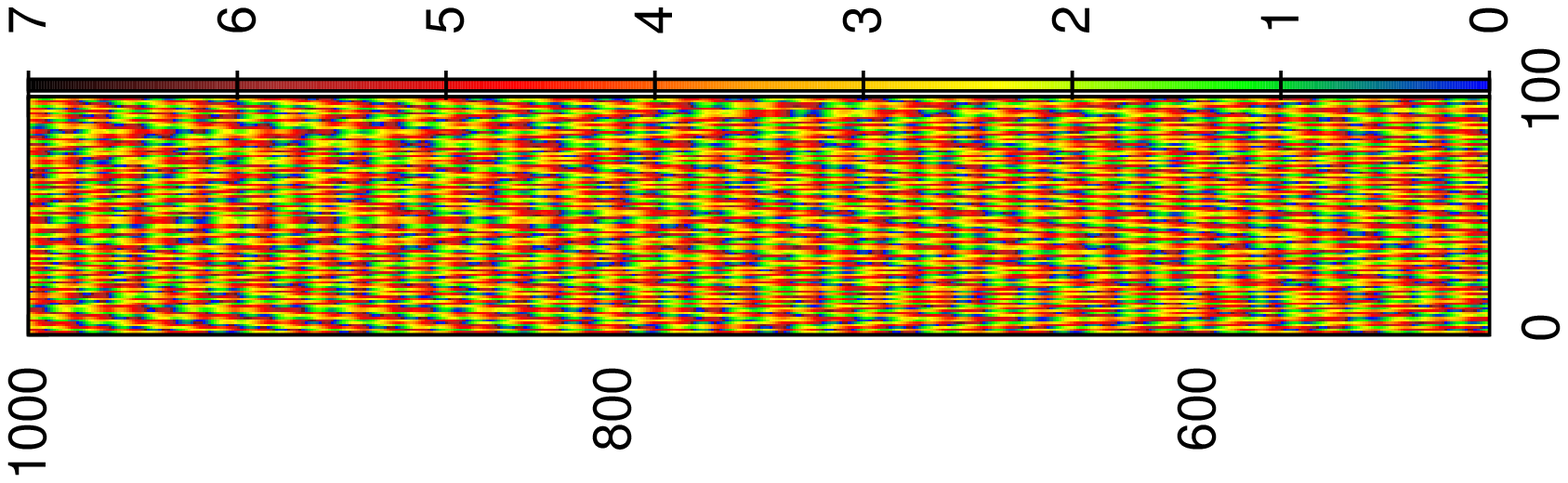}}
           \textbf{(e)}\resizebox{3 cm}{!}{\includegraphics[angle=-90]{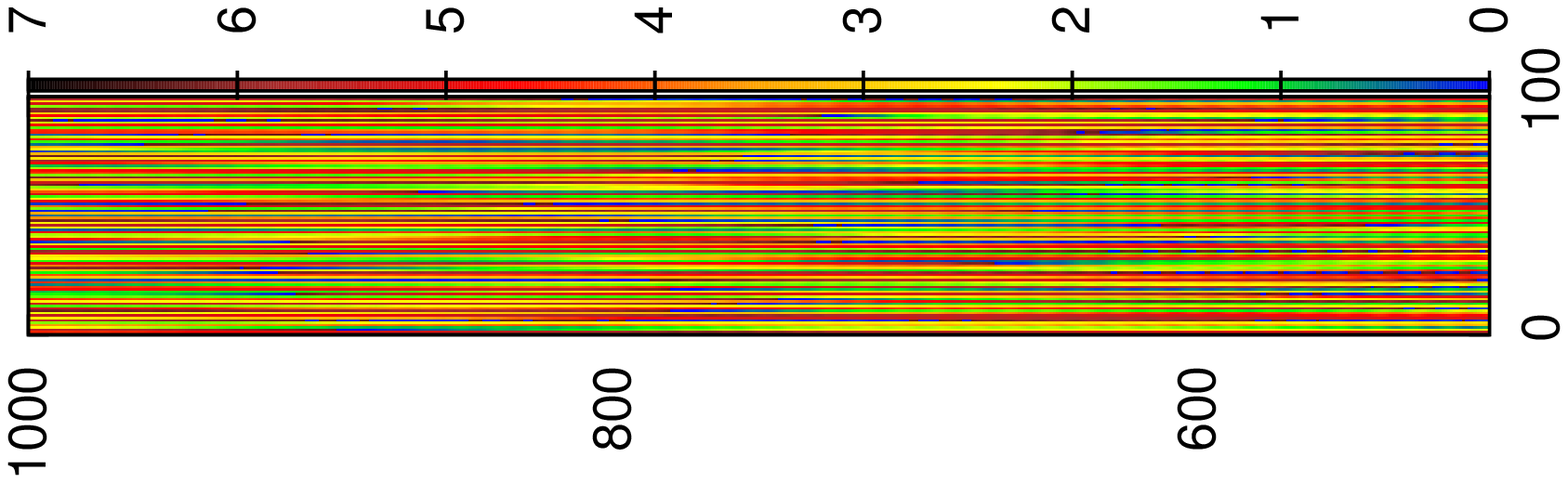}}\\
           \textbf{\large Oscillator Index}
                          \end{tabular}
 \caption{{(Color Online) Dynamics of chimera states in space and time with $N=100, r=0.35$ and $d=0.1$ between times $500$ to $1000$ unit for Gaussian quenched disorder with (a) $m=1.46$, (b) $m=1.55$, (c) $m=1.554$, (d) $m=1.555$ and (e) $m=1.56$ with identical initial random configuration of phases showing the progressive loss of order; the color bar is shown beside each case.}}
\end{center}
\end{figure}

\begin{figure}[h]
\begin{center}
\begin{tabular}{c}
        \textbf{(a)} \resizebox{8cm}{!}{\includegraphics[angle=-90]{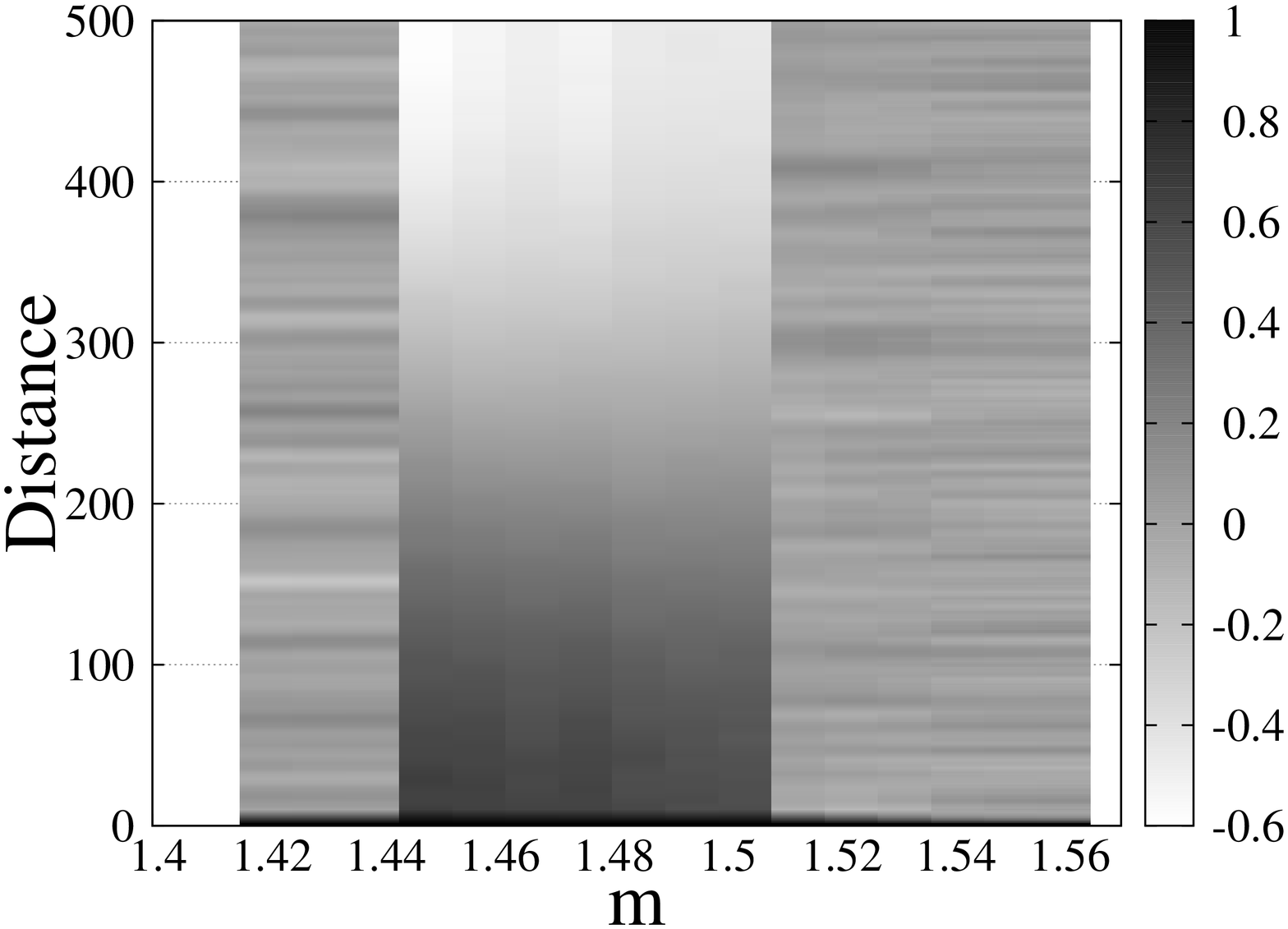}}
         \textbf{(b)}\resizebox{8cm}{!}{\includegraphics[angle=-90]{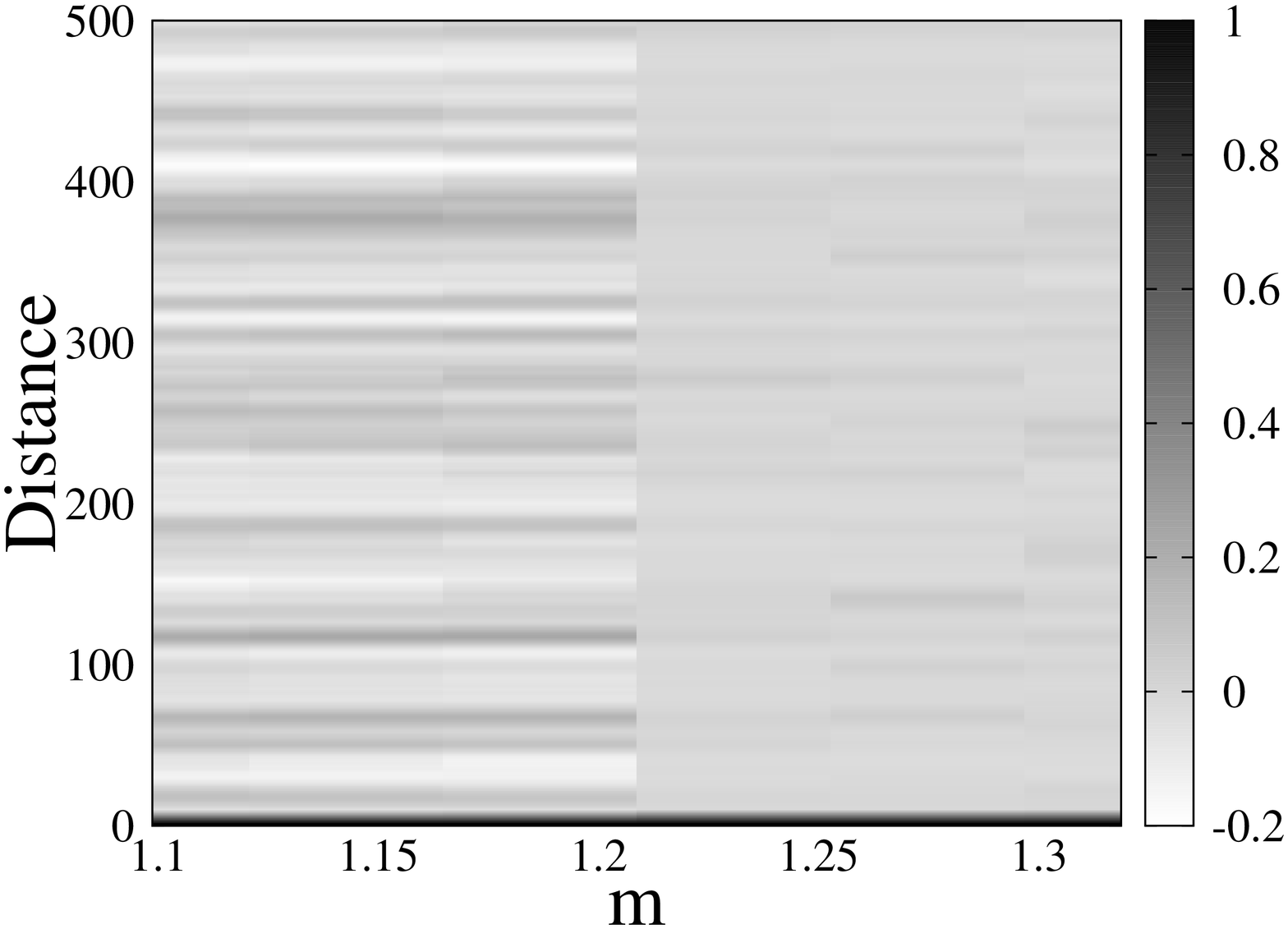}}\\
         \textbf{(c)}\resizebox{8cm}{!}{\includegraphics[angle=-90]{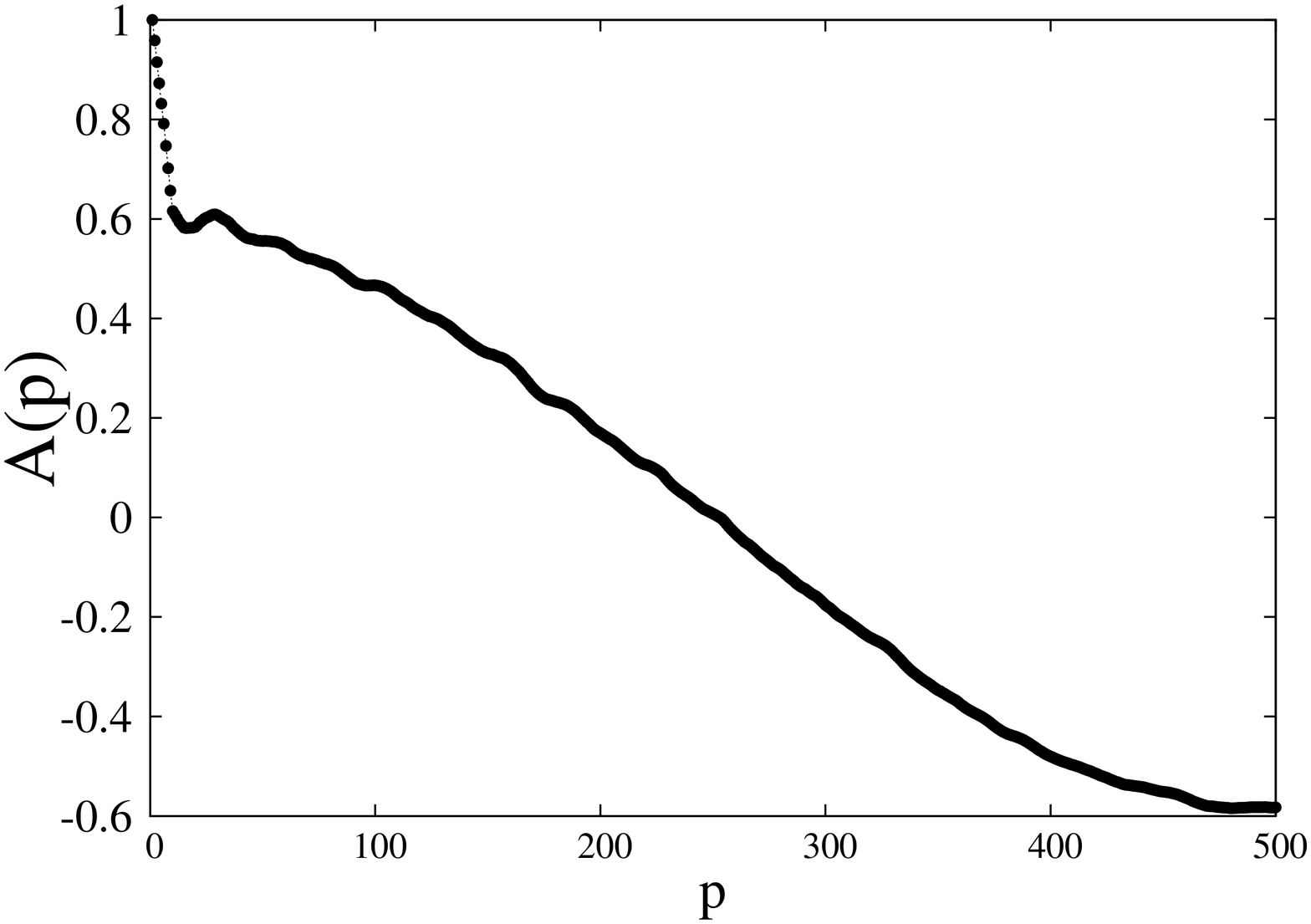}}
        
                  \end{tabular}
 \caption{{Normalized autocorrelation profile of $A(p)$ with distance $p$, after averaging between times 800 to 1000 unit with $R_{l}=4$ for quenched Gaussian disorder with $r=0.35,N=1000$ having (a) $d=0.2$ and (b) $d=0.85$ and the autocorrelation profile for a chimera state with (c)$m=1.445,d=0.2$.}}
\end{center}
\end{figure}
\newpage
\begin{figure}[h]
\begin{center}
\begin{tabular}{c}
        \textbf{(a)} \resizebox{3 cm}{!}{\includegraphics[angle=-90]{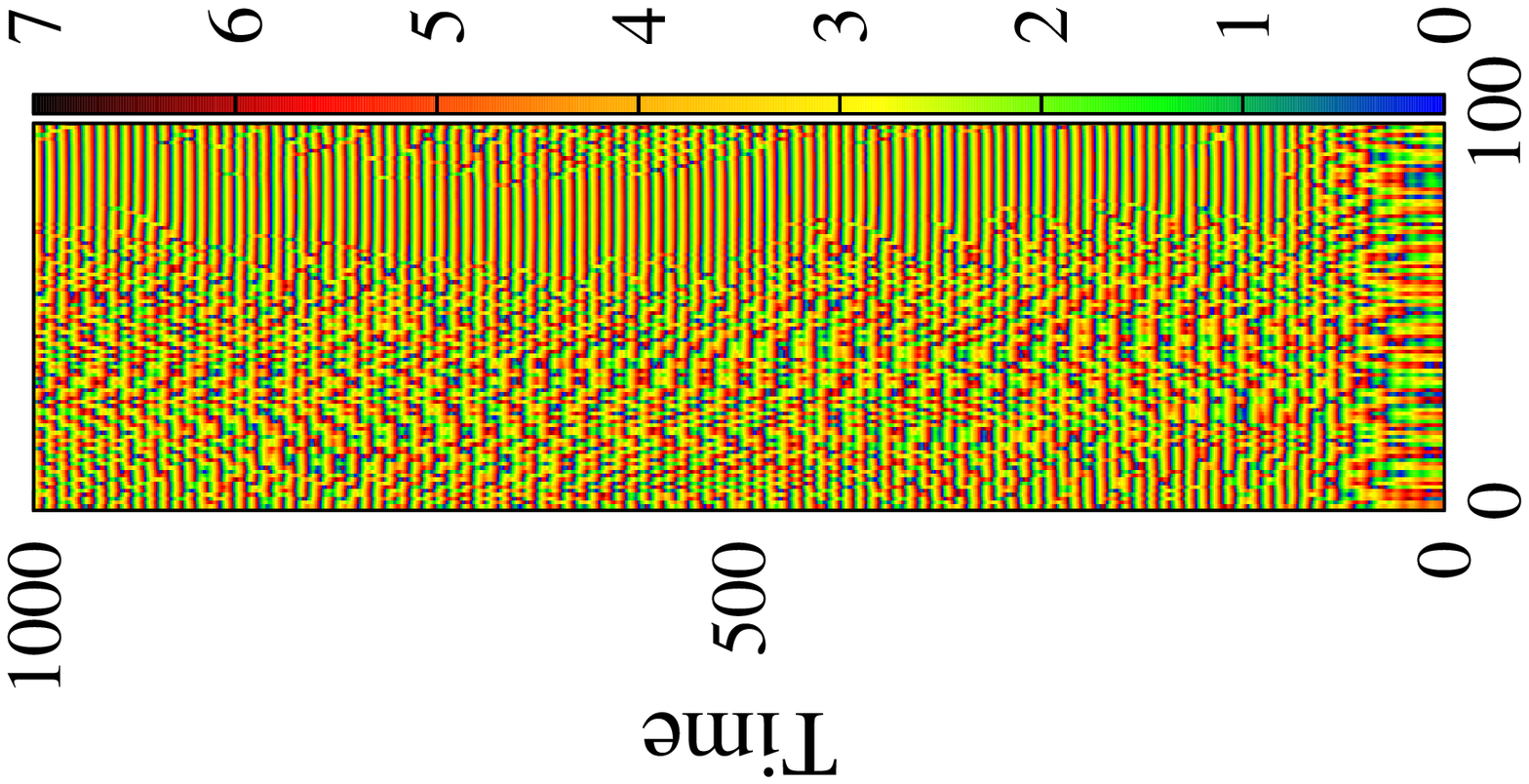}}
         \textbf{(b)}\resizebox{3 cm}{!}{\includegraphics[angle=-90]{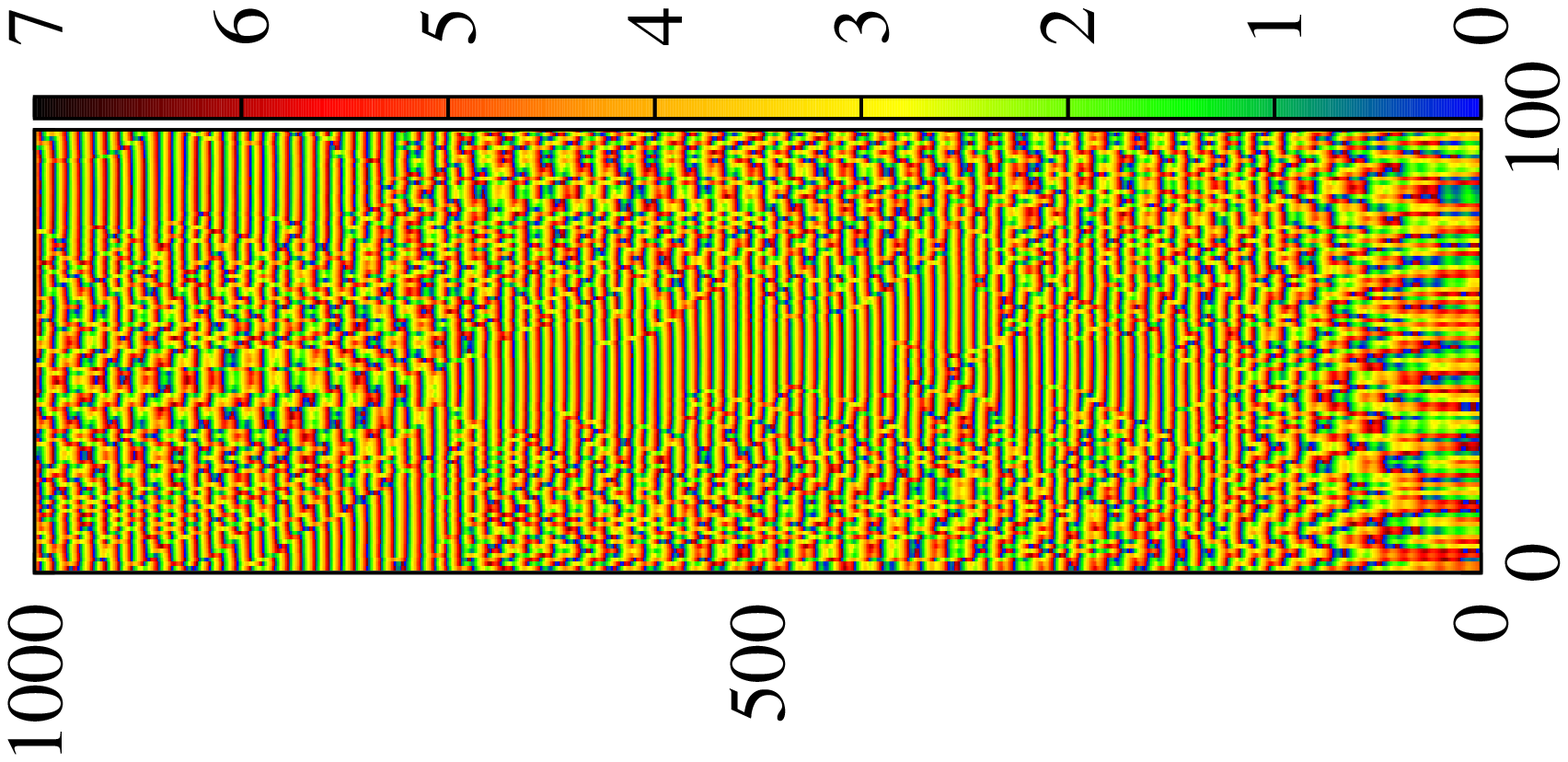}}
         \textbf{(c)}\resizebox{3 cm}{!}{\includegraphics[angle=-90]{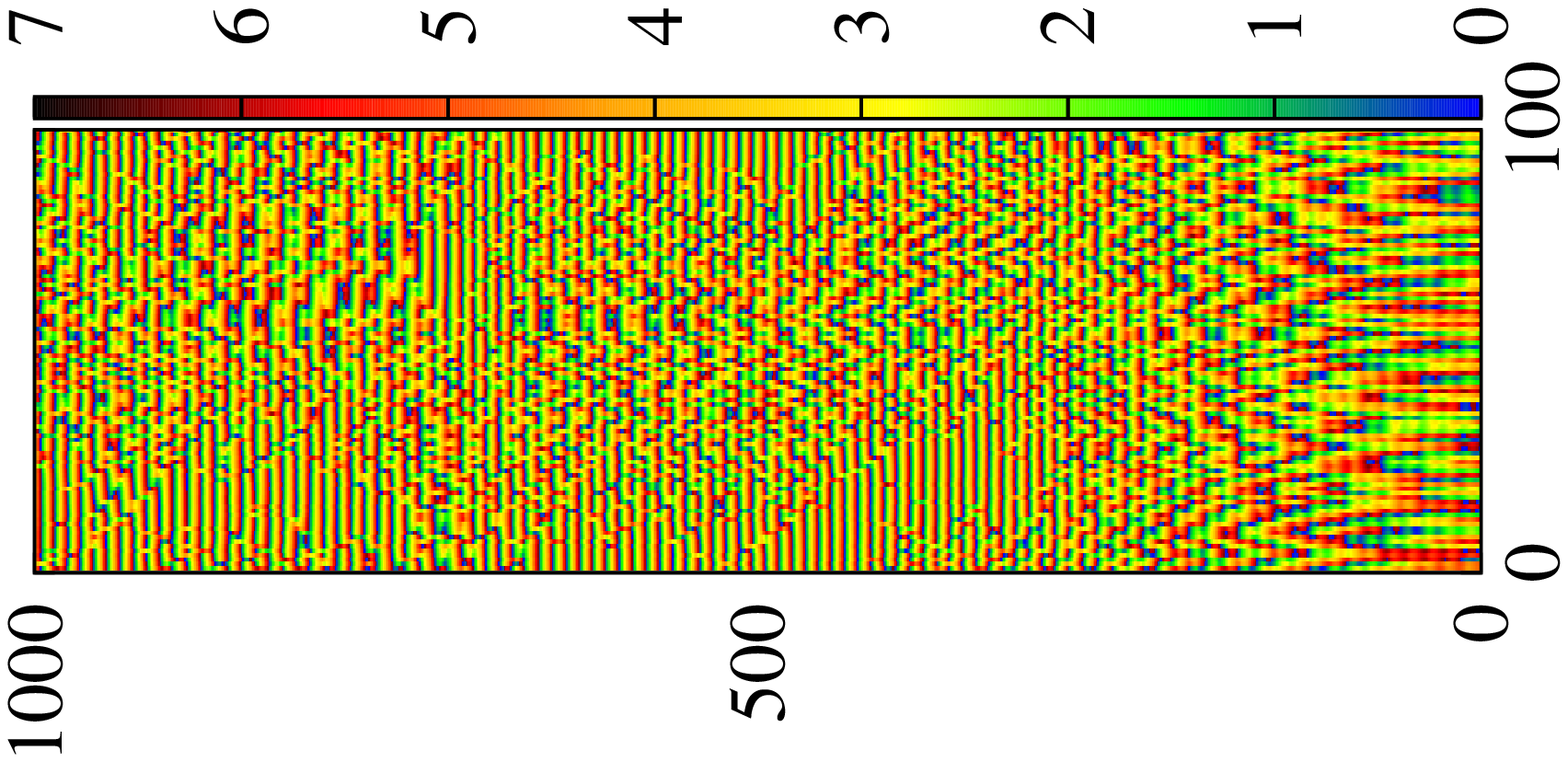}}\\
           \textbf{ Oscillator Index}
                          \end{tabular}
 \caption{{(Color Online) Dynamics of chimera states in space and time under annealed disorder with $N=100, r=0.35$ and $d=0.8$ with (a) $m=1.53$ (b) $m=1.55$ and (c) $m=1.56$.}}
\end{center}
\end{figure}
\newpage

\begin{figure}[h]
\begin{center}
\begin{tabular}{c}
        \resizebox{8cm}{!}{\includegraphics[angle=-90]{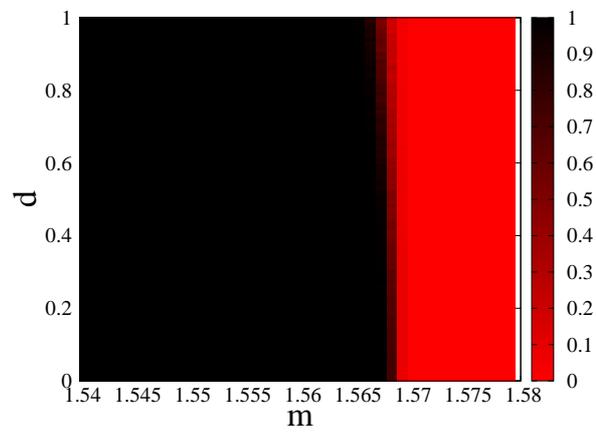}}

                  \end{tabular}
 \caption{{(Color Online) Phase Diagram for modulus of the order parameter $|Z|$ after time 5000 unit, obtained by simulation of the Ott-Antonsen equations with annealed disorder for $r=0.49$ and $N=100$; the values of $|Z|$ are each averaged over $200$ realizations of the time-series of $\alpha(t) $ for each set of values for $m$ and $d$.}}
\end{center}
\end{figure}

\end{document}